\documentclass[10pt]{article}
\usepackage[utf8]{inputenc}
\usepackage[a4paper,top=2.5cm,bottom=2cm,left=1.75cm,right=1.75cm,marginparwidth=2cm]{geometry}
\usepackage{amsmath}
\usepackage{amssymb}
\numberwithin{equation}{section}
\usepackage[english]{babel}
\usepackage{amsthm}
\usepackage{bbm}
\usepackage{amsfonts}
\usepackage{mathrsfs}
\usepackage{hyperref}
\usepackage{verbatim}
\usepackage{titlesec}
\usepackage{subcaption}
\usepackage{booktabs}
\usepackage{tikz}

\hypersetup{
    colorlinks=true,
    linkcolor=blue,
    citecolor=red,
    filecolor=magenta,      
    urlcolor=black,
    pdftitle={RegularizingDamping},
    pdfpagemode=FullScreen,
    }
\usepackage{graphicx, tikz}
\usepackage{float}

\usepackage{xcolor}
\usepackage{bbm}
\usepackage[format=plain,
            font=it]{caption}

\newtheorem{theorem}{Theorem}[section]

\newtheorem{lemma}[theorem]{Lemma}
\newtheorem{proposition}[theorem]{Proposition}
\newtheorem{remark}[theorem]{Remark}

\newtheorem{definition}[theorem]{Definition}

\def\G{\Gamma }

\def\k{\kappa }

\usepackage{todonotes}

\title{Regularization of singular Bloch spectra by damping in photonic crystals}

\author{Konstantinos Alexopoulos$^1$ \and Bryn Davies$^{2}$}
\date{}
\begin{document}

\maketitle

\vspace{-0.5cm}

{\footnotesize
$^1$ MAP5, Université Paris Cité, CNRS, Paris, France.

$^2$ Mathematics Institute, University of Warwick, Coventry CV4 7AL, UK.
}

\vspace{0.5cm}

\begin{abstract}
    The use of dispersive materials in photonic crystals leads to nonlinear spectral problems. The presence of resonances can lead to clustering of Bloch branches, however the effects of this intricate spectral structure are regularized by damping. We show that this regularization by damping is due to the displacement of branches into the complex-frequency plane, removing the resonant singularity from the real frequency axis. For a scalar photonic-crystal model with periodic subwavelength inclusions, we derive an interaction-matrix description that separates the material response from the geometry. Under spectral separation and continuation assumptions, these formulas quantify the damping-induced motion and its leading temporal decay rate. We also establish the local relation between complex-frequency branches and spatial attenuation at real frequency, distinguishing the continuation of a selected mode from the least-attenuated response. Two-dimensional plane-wave computations illustrate these complementary signatures of regularization. A reduced-solver benchmark demonstrates an approximately order-of-magnitude reduction in solve time for five selected branches at fixed geometry, with a quantified accuracy trade-off. The analysis thus provides both a physical interpretation of damping-induced spectral regularization and a computationally inexpensive description of selected resonant modes.
\end{abstract}

\section{Introduction}

Photonic crystals are periodic media whose band-gap structure can be used to control the propagation of electromagnetic waves. Since the foundational works on inhibited spontaneous emission and photon localization \cite{yablonovitch1987inhibited,john1987strong}, they have become a central example of wave control and band-gap phenomena \cite{joannopoulos2008molding,soukoulis2012photonic}. Applications include negative-refraction lenses~\cite{pendry2000negative}, cloaking by anomalous localized resonance~\cite{milton2006cloaking}, rainbow trapping~\cite{tsakmakidis2007trapped}, and topological photonic waveguides~\cite{khanikaev2013photonic}.

The mathematical description of periodic media such as photonic crystals is based on Floquet--Bloch theory \cite{kuchment2016overview}. The spectrum is obtained from quasiperiodic cell problems parametrized by the Bloch variable, a viewpoint originating in periodic quantum mechanics~\cite{kronig1931quantum} and now standard for periodic elliptic operators~\cite{kuchment2016overview}. Analogous methods apply to classical waves and structured media~\cite{adams2008bloch,movchan2002asymptotic}. Frequency-dependent material parameters increase the  difficulty of these spectral problems. Common models for the electromagnetic response of metals, such as Drude, Lorentz and Drude--Lorentz laws, contain both material resonances and damping \cite{maier2007plasmonics,schulkin2004analytical,li2003photonic,sehmi2020applying}. Resonances are described by singularities (typically, simple poles) in the permittivity, whereas damping is characterised by the permittivity having non-zero imaginary part. Consequently, the Bloch problem is nonlinear in frequency and the poles of the material law interact with the periodic geometry. Complex band structures and attenuation in dispersive and absorptive photonic crystals have been investigated both numerically and theoretically~\cite{engstrom2018spectral,brule2016calculation,xiao2021band}. The issue addressed here is the singular regime in which a material pole makes the contrast large enough to balance the small geometric scale of subwavelength inclusions.

Our first contribution is an effective description of that balance. Starting from a quasiperiodic Lippmann--Schwinger formulation and the subwavelength resonator framework of~\cite{ammari2022wave,ammari2024functional}, we isolate the material contrast from the compact geometric interaction operator. Its selected modes admit a finite-dimensional reduction. If $\lambda_j^\kappa(\delta,\omega)$ is a simple, uniformly controlled eigenvalue branch of this full interaction operator and $\omega^*$ is a real material pole with residue $r_*$, the associated undamped resonant branch satisfies
\begin{equation} \label{eq:intro-undamped-branch}
    \omega_j^0(\kappa)
    =
    \omega^*
    +
    \delta^2 C_*\lambda_j^\kappa(\delta,\omega^*)
    +
    O(\delta^4\Lambda_\delta^2),
    \qquad
    C_*:=\omega_*^2r_*,
\end{equation}
where $\Lambda_\delta=|\log\delta|/(2\pi)$ in two dimensions and $\Lambda_\delta=1$ in three dimensions. Thus, the real pole generates a family of dispersion curves. Their separation determines whether gaps occur, provided that the asymptotics remain uniform over the full Brillouin zone and no additional band enters the candidate interval.

Our other contribution is to show how damping continues this singular structure into the complex plane. In the one-dimensional problem, damping regularizes the gap hierarchy generated by real material singularities~\cite{alexopoulos2023effect}. Here, for the damped pole $\omega_+^\gamma$ and its residue $r_\gamma$, we obtain
\begin{equation} \label{eq:intro-damped-branch}
    \omega_j^\gamma(\kappa)
    =
    \omega_+^\gamma
    +
    \delta^2 C_\gamma\lambda_j^\kappa(\delta,\omega_+^\gamma)
    +
    O(\delta^4\Lambda_\delta^2),
    \qquad
    C_\gamma:=(\omega_+^\gamma)^2r_\gamma,
\end{equation}
and, in particular,
\begin{equation} \label{eq:intro-imaginary-displacement}
    \operatorname{Im}\omega_j^\gamma(\kappa)
    =
    -\frac{\gamma}{2\beta}
    +O(\delta^2\Lambda_\delta).
\end{equation}
Damping therefore moves the pole away from the real axis without eliminating the associated resonances. For real Bloch parameter they appear as temporally decaying complex frequencies. As a result, the singular nature of the Bloch spectra is still present. However, the damping has a regularising effect. 
This regularization is also reflected in the real-frequency description, where the Bloch parameter is allowed to become complex and its imaginary part describes spatial attenuation. Such complementary complex-frequency and complex-wavenumber descriptions of loss have been studied in \cite{laude2013effect}. In our numerical examples, the real-frequency attenuation profiles display a smoother response than the clustered complex-frequency spectra. This connects the analysis with complex-Bloch and generalized Brillouin-zone descriptions~\cite{ammari2023mathematical,brule2016calculation}.

The asymptotic reduction also has a computational motivation. Direct plane-wave discretizations \cite{johnson2001block} and finite-element formulations for dispersive photonic crystals \cite{brule2016calculation,xiao2021band} resolve the field throughout the periodic cell and require repeated spectral solves as the Bloch parameter or material damping varies. The asymptotic interaction matrices instead retain selected inclusion modes; with one mode per inclusion their dimension is $N$, independent of the number of degrees of freedom used to resolve the cell. Once the geometric interaction coefficients are available, these matrices offer a route to inexpensive evaluation of selected resonant branches near a material pole. This saving concerns the asymptotic approximation, not the exact Schur complement, whose complementary inverse can itself be costly. Alongside the full-solver spectral illustrations, we compare the runtimes and complex frequencies of the two numerical formulations at fixed geometry as damping varies. This benchmark measures the computational saving for five selected modes and their agreement with a finer plane-wave reference; it does not establish an equal-accuracy speedup, since the two solver configurations have different errors.

% The higher-dimensional geometry supplies an essential qualification rather than a separate mechanism. Complete gaps require separation over the entire Brillouin zone, whereas a high-symmetry path provides only a partial diagnostic. This is the issue emphasized in~\cite{craster2012dangers}. It is also consistent with high-frequency homogenization approaches for periodic dispersive media~\cite{doi:10.1137/23M159648X,craster2010high}, in which local dispersion-surface geometry determines effective wave behaviour.

The paper is organized as follows. Section~\ref{sec:problem-setting} introduces the periodic scalar model, the Drude--Lorentz law and the subwavelength scaling. Section~\ref{sec:bloch-band-gaps} formulates the Bloch problem and the notions of complete and partial gaps and attenuation. Section~\ref{sec:integral-formulation} gives the quasiperiodic Lippmann--Schwinger formulation, and Section~\ref{sec:subwavelength-asymptotics} derives the finite-dimensional reduction and the two- and three-dimensional asymptotics. Section~\ref{sec:material-singularities} establishes the branches generated by a real pole and the associated gap criterion. Section~\ref{sec:material-damping} proves their regularization under damping and relates temporal decay to spatial attenuation. Section~\ref{sec:numerical-investigations} presents the two-dimensional plane-wave illustrations. Operator-norm remainder estimates and numerical methods are collected in Appendices~\ref{app:kernel-remainders} and~\ref{app:numerical-methods}, respectively.

\section{Problem setting} \label{sec:problem-setting}

We consider a scalar periodic-wave model for a photonic crystal composed of small dispersive inclusions embedded in a homogeneous background medium. The physical dimension is denoted by $d \in \{2,3\}$. In two dimensions the model describes, for example, an appropriate transverse-magnetic polarization with constant permeability; in three dimensions it should be understood as a scalar analogue of the full Maxwell system, which is studied mostly as a mathematical toy problem. The inclusions are assumed to be non-magnetic, so that the magnetic permeability is constant and equal to $\mu_0$ throughout $\mathbb{R}^d$. The background medium has constant permittivity $\varepsilon_0>0$, while the inclusions have frequency-dependent permittivity.

\subsection{Dispersive material law}

The inclusions are modelled by a Drude--Lorentz type permittivity of the form
\begin{equation} \label{eq:permittivity}
    \varepsilon(\omega) = \varepsilon_0 + \frac{\alpha}{1-\beta \omega^2 - i\gamma \omega},
\end{equation}
where $\alpha,\beta,\gamma \geq 0$. The parameter $\alpha$ measures the strength of the dispersive response, $\beta$ determines the natural resonant frequency of the material, and $\gamma$ is the damping parameter. This model is motivated by dispersive optical materials, including halide perovskites.

The poles of $\varepsilon$ are given by
\begin{equation} \label{eq:material-poles}
    \omega_\pm^* = \frac{1}{2\beta} \left( -i\gamma \pm \sqrt{4\beta-\gamma^2} \right),
\end{equation}
whenever $\beta>0$. In the undamped case $\gamma=0$, these poles lie on the real axis and are located at $\omega_\pm^* = \pm 1/\sqrt{\beta}$. When $\gamma>0$, the poles move into the lower half of the complex frequency plane. This distinction will be central to the analysis below: real material poles produce singular spectral behaviour, whereas damping regularizes this singular structure.

Let us also introduce the contrast function $\xi(\omega) := \mu_0\Big(\varepsilon(\omega)-\varepsilon_0\Big)$. Then, for the material law \eqref{eq:permittivity}, we have that
\begin{equation} \label{eq:contrast-drude-lorentz}
    \xi(\omega) = \frac{\mu_0\alpha}{1-\beta\omega^2-i\gamma\omega}.
\end{equation}

\subsection{Dilute unit-cell geometry}

Let $D_1,\dots,D_N$ be $N\in\mathbb{N}$ bounded, simply connected reference domains in $\mathbb{R}^d$ with Lipschitz boundaries, and let $z_1,\dots,z_N$ be distinct points in the interior of the unit cell $Y$. We will work in the dilute subwavelength regime. For a characteristic size parameter $0<\delta\ll1$, the physical inclusions in one cell are
\begin{equation} \label{eq:scaled-cluster}
    D_j^\delta:=z_j+\delta D_j,
    \qquad
    \Omega_\delta:=\bigcup_{j=1}^N D_j^\delta .
\end{equation}
Thus the inclusion centres remain fixed as $\delta\to0$, while every component shrinks about its own centre. This is the fixed-cell dilute regime used throughout the analysis and in the numerical section. We assume that the centres remain uniformly separated from one another and from the cell boundary: there is a constant $c_0>0$, independent of $\delta$, such that
\begin{equation}
    \min_{i\neq j}|z_i-z_j|\geq c_0,
    \qquad
    \operatorname{dist}(z_j,\partial Y)\geq c_0.
\end{equation}
For sufficiently small $\delta$, the sets $D_j^\delta$ are therefore pairwise disjoint and remain separated from their periodic copies.

\subsection{Periodic structure}

Let $\Lambda$ be a lattice in $\mathbb{R}^d$ generated by linearly independent vectors $l_1,\dots,l_d \in \mathbb{R}^d$ so that
\begin{equation} \label{eq:lattice}
    \Lambda
    :=
    \left\{
        m_1l_1+\cdots+m_dl_d
        \;:\;
        m_1,\dots,m_d\in\mathbb{Z}
    \right\}.
\end{equation}
The corresponding fundamental cell is $Y
    :=
    \left\{
        c_1l_1+\cdots+c_dl_d
        \;:\;
        0\leq c_1,\dots,c_d<1
    \right\}$. 
The dual lattice $\Lambda^*$ is generated by vectors $\alpha_1,\dots,\alpha_d$ satisfying $\alpha_i\cdot l_j = 2\pi\delta_{ij}$ for $i,j=1,\dots,d$. The Brillouin zone is denoted by $Y^* := \mathbb{R}^d/\Lambda^*$ and the full periodic array of resonators is
\begin{equation} \label{eq:periodic-structure}
    \mathcal{D}
    :=
    \bigcup_{m\in\Lambda}
    \bigcup_{j=1}^N
    \Big(D_j^\delta+m\Big).
\end{equation}
Inside $\mathcal{D}$ the permittivity is $\varepsilon(\omega)$, while outside $\mathcal{D}$ it is equal to the background value $\varepsilon_0$.

\begin{figure}
    \centering
    \includegraphics[width=0.65\linewidth]{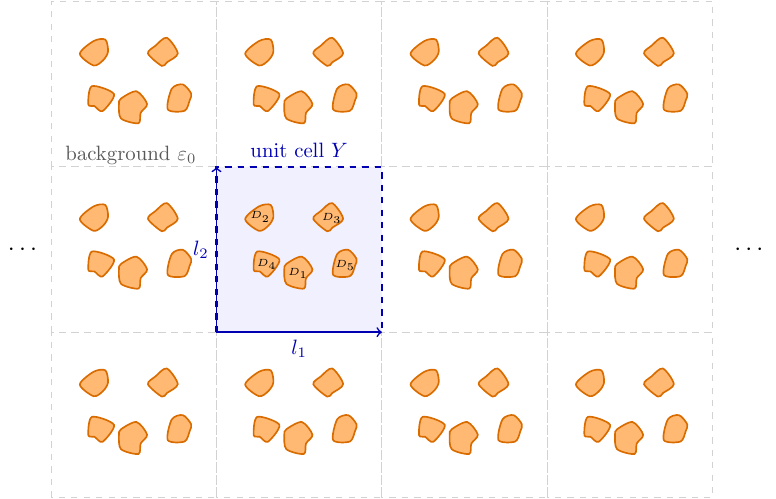}
    \caption{Schematic depiction of a two-dimensional periodic array of subwavelength dispersive inclusions. Each unit cell contains five resonators $D_1,\dots,D_5$, scaled by the small parameter $\delta$, embedded in a homogeneous background medium with permittivity $\varepsilon_0$.}
    \label{fig:periodic-2d-schematic}
\end{figure}

\subsection{Helmholtz formulation}

We study time-harmonic scalar waves with angular frequency $\omega$. This scalar model corresponds, for instance, to transverse magnetic polarized electromagnetic waves in an appropriate two-dimensional reduction. The field $u$ satisfies
\begin{equation} \label{eq:periodic-helmholtz}
    \begin{cases}
        \Delta u + \omega^2 \varepsilon(\omega)\mu_0 u = 0,
        & \text{in } \mathcal{D},\\[2mm]
        \Delta u + k_0^2 u = 0,
        & \text{in } \mathbb{R}^d\setminus\overline{\mathcal{D}},\\[2mm]
        u|_+ - u|_- = 0,
        & \text{on } \partial\mathcal{D},\\[2mm]
        \displaystyle
        \frac{\partial u}{\partial \nu}\bigg|_+
        -
        \frac{\partial u}{\partial \nu}\bigg|_-
        =0,
        & \text{on } \partial\mathcal{D},
    \end{cases}
\end{equation}
where $k_0 := \omega\sqrt{\varepsilon_0\mu_0}$ is the background wavenumber, and $\nu$ denotes the outward unit normal to the inclusions. The symbols $|_+$ and $|_-$ denote traces taken from the exterior and interior of the inclusions, respectively.

\begin{remark}
    The asymptotic regime studied here is performed by rescaling each inclusion separately by writing $x=z_i+\delta X$ on $D_i^\delta$. The lattice and the physical Bloch parameter are not rescaled. The resulting kernels therefore have the form
    \begin{equation}
        G^{\kappa,k_0}\big(z_i-z_j+\delta(X-Y)\big),
    \end{equation}
    rather than a quasiperiodic Green function with the same lattice and wavenumber $\delta k_0$. Factoring the local singularity of this kernel produces the common prefactor $\delta^2$ in both dimensions: in two dimensions it is accompanied by $|\log\delta|$, whereas in three dimensions it multiplies the Newtonian potential.
\end{remark}

\section{Bloch formulation and band-gap notions}
\label{sec:bloch-band-gaps}

We now pass from the full periodic problem to a family of quasiperiodic cell problems by means of the Floquet--Bloch transform. This formulation is the natural starting point for defining Bloch spectra, spectral bands and band gaps. It is also the framework in which the effect of damping can be interpreted through complex quasiperiodicities.

\subsection{Floquet--Bloch decomposition}

\begin{definition}
    A function $u:\mathbb{R}^d\to\mathbb{C}$ is said to be $\kappa$-quasiperiodic with respect to the lattice $\Lambda$ if
    \begin{equation}
        u(x+m)=e^{i\kappa\cdot m}u(x), \qquad x\in\mathbb{R}^d,\quad m\in\Lambda,
    \end{equation}
    or, equivalently, if $e^{-i\kappa\cdot x}u(x)$ is $\Lambda$-periodic. Here $\kappa\in Y^*$ is the Bloch parameter.
\end{definition}

\begin{definition}
    For sufficiently regular functions, the \emph{Floquet--Bloch transform} is defined by
    \begin{equation}
        \mathcal{F}[u](x,\kappa) := \sum_{m\in\Lambda} u(x-m)e^{i\kappa\cdot m}, \qquad x\in Y,\quad \kappa\in Y^*.
    \end{equation}
    Reindexing with $m=n+\ell$ gives
    \begin{equation}
        \mathcal F[u](x+n,\kappa)
        =\sum_{\ell\in\Lambda}u(x-\ell)e^{i\kappa\cdot(n+\ell)}
        =e^{i\kappa\cdot n}\mathcal F[u](x,\kappa).
    \end{equation}
    Thus the transform has the stated $\kappa$-quasiperiodicity. Conversely, the inverse transform is given by
    \begin{equation}
        u(x) = \frac{1}{|Y^*|} \int_{Y^*} \mathcal{F}[u](x,\kappa) \mathrm{d}\kappa,
    \end{equation}
    where $\mathcal{F}[u](\cdot,\kappa)$ is extended quasiperiodically outside the unit cell.
\end{definition}

For each $\kappa\in Y^*$, we denote by
\begin{equation}
    u_{\kappa}(x) := \mathcal{F}[u](x,\kappa)
\end{equation}
the $\kappa$-quasiperiodic Bloch component of the field $u$. Applying the Floquet--Bloch transform to the periodic Helmholtz problem \eqref{eq:periodic-helmholtz}, we obtain, for each $\kappa\in Y^*$, the quasiperiodic transmission problem
\begin{equation} \label{eq:quasiperiodic-helmholtz}
    \begin{cases}
        \Delta u_\kappa + \omega^2\varepsilon(\omega)\mu_0 u_\kappa = 0,
        & \text{in } \mathcal{D},\\[2mm]
        \Delta u_\kappa + k_0^2 u_\kappa = 0,
        & \text{in } \mathbb{R}^d\setminus\overline{\mathcal{D}},\\[2mm]
        u_\kappa|_+ - u_\kappa|_- = 0,
        & \text{on } \partial\mathcal{D},\\[2mm]
        \displaystyle
        \frac{\partial u_\kappa}{\partial \nu}\bigg|_+
        -
        \frac{\partial u_\kappa}{\partial \nu}\bigg|_-
        =0,
        & \text{on } \partial\mathcal{D},\\[2mm]
        u_\kappa(x+m)=e^{i\kappa\cdot m}u_\kappa(x),
        & m\in\Lambda .
    \end{cases}
\end{equation}
Equivalently, after restricting to one fundamental cell, one may impose quasiperiodic boundary conditions on $\partial Y$.

\begin{definition}
    The Bloch spectrum is the set of real frequencies $\omega$ for which \eqref{eq:quasiperiodic-helmholtz} admits a non-trivial solution for some real Bloch parameter $\kappa\in Y^*$.    
\end{definition}

\begin{remark}
    Because the permittivity depends on $\omega$, this is a nonlinear spectral problem: for each fixed $\kappa$, one seeks non-trivial solutions of
\begin{equation} \label{eq:nonlinear-bloch-problem}
    L_\kappa(\omega)u_\kappa=0.
\end{equation}
Here, for frequencies away from the material poles, the operator pencil is
\begin{equation} \label{eq:bloch-pencil-definition}
    L_\kappa(\omega):H^2_\kappa(Y)\longrightarrow L^2(Y),
    \qquad L_\kappa(\omega)v=-\Delta v-\mu_0\omega^2\varepsilon_Y(x,\omega)v,
\end{equation}
where
\begin{equation}
    \varepsilon_Y(x,\omega)
    :=\varepsilon_0+(\varepsilon(\omega)-\varepsilon_0)
    \mathbf 1_{\mathcal D\cap Y}(x),
    \qquad
    H^2_\kappa(Y):=\{e^{i\kappa\cdot x}p(x):p\in H^2_{\mathrm{per}}(Y)\}.
\end{equation}
The space $H^2_{\mathrm{per}}(Y)$ is the periodic Sobolev space on the cell; thus the domain encodes quasiperiodicity of the field and its first derivatives. The distributional equation $L_\kappa(\omega)v=0$ incorporates the transmission conditions in \eqref{eq:quasiperiodic-helmholtz}. Its dependence on $\omega$ is nonlinear through both the factor $\omega^2$ and the material law.
\end{remark}

\subsection{Bloch bands and band gaps}

In the absence of damping, the permittivity is real-valued for real frequencies away from its poles. In this case, the solutions of \eqref{eq:quasiperiodic-helmholtz} give rise to Bloch branches
\begin{equation}
    \omega_n:Y^*\to\mathbb{R}, \qquad n=1,2,\dots,
\end{equation}
where $\omega_n(\kappa)$ denotes the $n$th frequency associated with the Bloch parameter $\kappa$. The corresponding spectral bands are the sets
\begin{equation}
    B_n := \Big\{ \omega_n(\kappa): \kappa\in Y^* \Big\},
\end{equation}
and the spectrum of the differential operator is given by the union of all the spectral bands $\cup_{n\geq 1} B_n$ \cite{kuchment2016overview}.

For undamped systems, the graph of $\omega_n$ over $Y^*$ is a dispersion surface whose image $B_n\subset\mathbb{R}$ is a closed interval. In one dimension this interval is easy to calculate since $Y^*$ is an interval. In more than one dimension it is common practice to check only the values of $\kappa$ that fall on a path connecting the high-symmetry points. This often reveals the extrema allowing band gaps to be identified. In what follows, the term band gap will refer to a complete band gap unless stated otherwise. We will also use the weaker language of partial gaps when discussing gaps observed only along selected curves in the Brillouin zone.

\subsection{Complex quasiperiodicity and attenuation}

When damping is present, the permittivity $\varepsilon(\omega)$ is complex-valued for real $\omega$. The corresponding Bloch problem is no longer self-adjoint, and one must choose how to encode attenuation. There are two common conventions. One may keep $\kappa$ real and allow $\omega$ to be complex, in which case $\operatorname{Im}\omega$ describes temporal decay. Alternatively, one may keep $\omega$ real and allow $\kappa$ to be complex, in which case $\operatorname{Im}\kappa$ describes spatial decay. We use both parametrizations: the first exposes the motion of the resonant spectrum under damping, while the second describes the real-frequency attenuation observed across the lattice.

Thus we write
\begin{equation}
    \kappa = \kappa_r + i\kappa_i,
    \qquad
    \kappa_r\in Y^*,
\end{equation}
where $\kappa_i$ measures spatial attenuation. In this setting, a mode behaves as
\begin{equation}
    u_\kappa(x+m)
    =
    e^{i\kappa_r\cdot m}
    e^{-\kappa_i\cdot m}
    u_\kappa(x),
    \qquad m\in\Lambda.
\end{equation}
Hence non-zero $\kappa_i$ corresponds to exponential growth or decay across the lattice, depending on the direction of propagation.

For complex material parameters, it is no longer appropriate to define a band gap simply as an interval in which no real Bloch mode exists, since material damping already produces attenuation. Instead, the relevant question is whether the periodic geometry produces additional attenuation beyond the background material loss.

\begin{definition}[Attenuation region]
Let $\omega\in\mathbb{R}$ and suppose that the quasiperiodic problem admits a solution with complex Bloch parameter $\kappa(\omega)$. We say that a frequency interval $I$ is an attenuation region if $|\operatorname{Im}\kappa(\omega)|$ is locally enhanced for $\omega\in I$ relative to neighbouring frequencies.
\end{definition}

\begin{remark}
    This definition is intentionally qualitative at this stage. In the one-dimensional problem \cite{alexopoulos2023effect}, the imaginary part of the Bloch parameter can be computed explicitly from the dispersion relation, and attenuation regions can be identified as local maxima of $|\operatorname{Im}\kappa|$. In the multidimensional problem, the analogous notion depends on the direction in the Brillouin zone and on the choice of complex continuation of $\kappa$.    
\end{remark}

\section{Quasiperiodic integral formulation} \label{sec:integral-formulation}

We now reformulate the quasiperiodic Helmholtz problem on the fixed physical cell and then rescale each inclusion separately. This distinction is important: the lattice and the Bloch parameter remain fixed as $\delta\to0$, while the arguments of the quasiperiodic Green function approach either the origin or the separation vector between two inclusion centres.

\subsection{Quasiperiodic Green's function}

Let $G^k$ denote the outgoing Green's function for the Helmholtz operator in $\mathbb{R}^d$, namely
\begin{equation}
    (\Delta + k^2)G^k = \delta_0 \qquad \text{in } \mathbb{R}^d,
\end{equation}
together with the outgoing radiation condition. Here $\delta_0$ denotes the Dirac mass at the origin. In dimensions $d=2,3$, it is given by
\begin{equation} \label{eq:free-green}
    G^k(x) =
    \begin{cases}
        \displaystyle
        -\frac{i}{4}H_0^{(1)}(k|x|), & d=2,\\[3mm]
        \displaystyle
        -\frac{e^{ik|x|}}{4\pi |x|}, & d=3,
    \end{cases}
\end{equation}
where $H_0^{(1)}$ is the Hankel function of the first kind and order zero.

For $\kappa\in Y^*$ and away from the diffraction set
\begin{equation} \label{eq:diffraction-set}
    \mathscr D(k):=
    \left\{\kappa\in Y^*:k^2=|\kappa+q|^2
    \text{ for some }q\in\Lambda^*\right\},
\end{equation}
the $\kappa$-quasiperiodic Green function is defined by the reciprocal-lattice representation
\begin{equation} \label{eq:quasiperiodic-green}
    G^{\kappa,k}(x)
    :=
    \frac{1}{|Y|}
    \sum_{q\in\Lambda^*}
    \frac{e^{i(\kappa+q)\cdot x}}{k^2-|\kappa+q|^2}.
\end{equation}
The series is understood in the standard distributional, limiting-absorption or Ewald sense. It is the unique quasiperiodic solution of
\begin{equation}
    (\Delta+k^2)G^{\kappa,k}
    =
    \sum_{m\in\Lambda}e^{i\kappa\cdot m}\delta_m
\end{equation}
and satisfies $G^{\kappa,k}(x+m) = e^{i\kappa\cdot m}G^{\kappa,k}(x)$, for $m\in\Lambda$. The direct real-space lattice sum is only conditionally convergent.

Locally, the quasiperiodic Green function has the same singularity as the free-space fundamental solution. More precisely,
\begin{equation} \label{eq:local-green-decomposition}
    G^{\kappa,k}(x)=
    \begin{cases}
        \displaystyle \frac{1}{2\pi}\log|x|+R^{\kappa,k}_2(x),&d=2,\\[2mm]
        \displaystyle -\frac{1}{4\pi|x|}+R^{\kappa,k}_3(x),&d=3,
    \end{cases}
\end{equation}
where the remainders $R^{\kappa,k}_d$ extend continuously to the origin, but are not generally smooth there. The smooth local remainder is obtained by subtracting the full Helmholtz fundamental solution:
\begin{equation} \label{eq:smooth-helmholtz-remainder}
    H_d^{\kappa,k}(x):=G^{\kappa,k}(x)-G^k(x).
\end{equation}
In a ball containing no other lattice point, the two Dirac singularities cancel, so
\begin{equation}
    (\Delta+k^2)H_d^{\kappa,k}=0.
\end{equation}
Interior elliptic regularity therefore gives smoothness of $H_d^{\kappa,k}$ in $x$; see~\cite{mclean2000strongly}.

More explicitly, for $r=|x|\to0$, the free-space formulas yield
\begin{align} \label{eq:laplace-subtracted-remainders}
    R_2^{\kappa,k}(x)
    &=H_2^{\kappa,k}(x)+c_2(k)
    +O\big(r^2(1+|\log r|)\big),\\
    R_3^{\kappa,k}(x)
    &=H_3^{\kappa,k}(x)-\frac{ik}{4\pi}
    +\frac{k^2r}{8\pi}+O(r^2),
\end{align}
where
\begin{equation}
    c_2(k):=\frac{\log(k/2)+\gamma_{\mathrm E}}{2\pi}-\frac{i}{4},
\end{equation}
and $\gamma_{\mathrm E}$ is Euler's constant. In particular, $R_2$ is locally $C^1$ but need not be $C^2$, while $R_3$ is locally Lipschitz but need not be differentiable at the origin. Both satisfy the estimate needed below,
\begin{equation} \label{eq:green-remainder-lipschitz}
    |R_d^{\kappa,k}(x)-R_d^{\kappa,k}(0)|\leq C|x|,
    \qquad d=2,3,
\end{equation}
uniformly on the specified compact parameter sets.

\subsection{Lippmann--Schwinger representation}

The differential formulation can be reformulated as an integral equation over the physical inclusions. Rescaling each inclusion to its own fixed reference domain then gives the exact operator family used below.

\begin{lemma}[Quasiperiodic Lippmann--Schwinger representation]
\label{thm:lippmann-schwinger-representation}
Let
\begin{equation}
    U_{\kappa,i}(X):=u_\kappa(z_i+\delta X),
    \qquad X\in D_i,
\end{equation}
and define $U_{\kappa,i}^{\mathrm{in}}$ similarly. The quasiperiodic Helmholtz problem is equivalent, on the inclusions, to
\begin{equation} \label{eq:lippmann-schwinger}
    U_{\kappa,i}(X)-U_{\kappa,i}^{\mathrm{in}}(X)
    =
    \delta^2\omega^2\xi(\omega)
    \sum_{j=1}^N
    \mathcal K_{ij}^{\kappa,\omega,\delta}
    [U_{\kappa,j}](X),
    \qquad X\in D_i,
\end{equation}
where
\begin{equation} \label{eq:scaled-block-operator}
    \mathcal K_{ij}^{\kappa,\omega,\delta}[\varphi](X)
    :=
    -\delta^{d-2}
    \int_{D_j}
    G^{\kappa,k_0}
    \big(z_i-z_j+\delta(X-Y)\big)
    \varphi(Y)\,\mathrm dY.
\end{equation}
\end{lemma}

\begin{proof}
Since $\omega^2\mu_0\varepsilon(\omega)-k_0^2=\omega^2\xi(\omega)$, the transmission problem can be written distributionally as
\begin{equation}
    (\Delta+k_0^2)u_\kappa
    =-\omega^2\xi(\omega)\mathbf 1_{\mathcal D}u_\kappa.
\end{equation}
Using the $\kappa$-quasiperiodic Green function and a homogeneous incident field
$(\Delta+k_0^2)u_\kappa^{\mathrm{in}}=0$ gives
\begin{equation}
    u_\kappa(x)-u_\kappa^{\mathrm{in}}(x)
    =
    -\omega^2\xi(\omega)
    \sum_{j=1}^N
    \int_{D_j^\delta}
    G^{\kappa,k_0}(x-y)u_\kappa(y)\,\mathrm dy.
\end{equation}
For $x=z_i+\delta X$ and $y=z_j+\delta Y$, one has $\mathrm dy=\delta^d\,\mathrm dY$ and $x-y=z_i-z_j+\delta(X-Y)$ meaning that
\begin{align}
    U_{\kappa,i}(X)-U_{\kappa,i}^{\mathrm{in}}(X)
    &=-\omega^2\xi(\omega)\delta^d
    \sum_{j=1}^N\int_{D_j}
    G^{\kappa,k_0}\big(z_i-z_j+\delta(X-Y)\big)
    U_{\kappa,j}(Y)\,\mathrm dY\notag\\
    &=\delta^2\omega^2\xi(\omega)
    \sum_{j=1}^N
    \mathcal K_{ij}^{\kappa,\omega,\delta}[U_{\kappa,j}](X),
\end{align}
which is \eqref{eq:lippmann-schwinger}.

Conversely, undoing this change of variables and applying $\Delta+k_0^2$ gives
\begin{align}
    (\Delta+k_0^2)u_\kappa
    &=(\Delta+k_0^2)u_\kappa^{\mathrm{in}}
    -\omega^2\xi(\omega)
    \sum_{j=1}^N\int_{D_j^\delta}
    (\Delta+k_0^2)G^{\kappa,k_0}(x-y)u_\kappa(y)\,\mathrm dy\notag\\
    &=-\omega^2\xi(\omega)\mathbf 1_{\mathcal D}(x)u_\kappa(x).
\end{align}
The volume potential and its first derivatives have matching traces across each inclusion boundary, so $u_\kappa$ and its normal derivative are continuous. Hence the integral equation recovers the transmission problem; see also~\cite{AD,alexopoulos2022mathematical}.
\end{proof}

The spectral problem for finding the Bloch bands has the incident field equal to zero. Thus, the characteristic frequencies are those values of $\omega$ for which there exists a non-trivial solution of
\begin{equation} \label{eq:lippmann-schwinger-resonance}
    U_\kappa
    =
    \delta^2\omega^2\xi(\omega)
    \mathcal K_\kappa(\omega,\delta)U_\kappa,
\end{equation}
where $\mathcal K_\kappa(\omega,\delta)$ is the block operator with entries \eqref{eq:scaled-block-operator} on
\begin{equation}
    \mathcal H:=\bigoplus_{j=1}^N L^2(D_j).
\end{equation}

\subsection{Integral operators and resonance condition}

Set
\begin{equation} \label{eq:operator-pencil-corrected}
    \tau(\omega,\delta):=\delta^2\omega^2\xi(\omega),
    \qquad
    \mathcal A_\kappa(\omega,\delta)
    :=I_{\mathcal H}-\tau(\omega,\delta)\mathcal K_\kappa(\omega,\delta).
\end{equation}
For fixed $\kappa$, the Bloch spectral problem is to find $\omega\in\mathbb C$ and $0\neq U_\kappa\in\mathcal H$ such that $\mathcal{A}_\kappa(\omega,\delta)U_\kappa=0$. Equivalently, the resonant frequencies are characterized by the nonlinear spectral condition
\begin{equation} \label{eq:operator-resonance-condition}
    \ker \mathcal{A}_\kappa(\omega,\delta)\neq\{0\}.
\end{equation}
The values of $\omega$ satisfying \eqref{eq:operator-resonance-condition} are the characteristic values of the quasiperiodic operator pencil $\mathcal{A}_\kappa(\cdot,\delta)$. This is the operator-valued analogue of a dispersion relation.

The local decomposition \eqref{eq:local-green-decomposition} gives the following operator-norm expansions, uniformly for $\kappa$ in compact subsets away from $\mathscr D(k_0)$. In two dimensions,
\begin{align}
    \mathcal K_{ii}^{\kappa,\omega,\delta}
    &=-\frac{\log\delta}{2\pi}\mathcal P_i+\mathcal K_{ii}^{\kappa,\omega,(0)}+O(\delta),
    \label{eq:correct-2d-self}\\
    \mathcal K_{ij}^{\kappa,\omega,\delta}
    &=\mathcal K_{ij}^{\kappa,\omega,(0)}+O(\delta),\qquad i\neq j,
    \label{eq:correct-2d-cross}
\end{align}
where
\begin{equation}
    % L_\delta:=-\frac{\log\delta}{2\pi},
    % \qquad
    \mathcal P_i[\varphi]:=\left(\int_{D_i}\varphi\right)\mathbf 1_{D_i},
\end{equation}
and
\begin{align}
    \mathcal K_{ii}^{\kappa,\omega,(0)}[\varphi](X)
    &:=-\int_{D_i}
    \left(\frac{1}{2\pi}\log|X-Y|+R_2^{\kappa,k_0}(0)\right)
    \varphi(Y)\,\mathrm dY,\\
    \mathcal K_{ij}^{\kappa,\omega,(0)}[\varphi](X)
    &:=-G^{\kappa,k_0}(z_i-z_j)\int_{D_j}\varphi(Y)\,\mathrm dY,
    \qquad i\neq j.
\end{align}
In three dimensions,
\begin{align}
    \mathcal K_{ii}^{\kappa,\omega,\delta}
    &=\mathcal N_i+\delta\mathcal K_{ii}^{\kappa,\omega,(1)}+O(\delta^2),
    \label{eq:correct-3d-self}\\
    \mathcal K_{ij}^{\kappa,\omega,\delta}
    &=\delta\mathcal K_{ij}^{\kappa,\omega,(1)}+O(\delta^2),\qquad i\neq j,
    \label{eq:correct-3d-cross}
\end{align}
where
\begin{equation}
    \mathcal N_i[\varphi](X)
    :=\frac{1}{4\pi}\int_{D_i}\frac{\varphi(Y)}{|X-Y|}\,\mathrm dY,
\end{equation}
and the first regular corrections are obtained from $R_3^{\kappa,k_0}(0)$ and the values $G^{\kappa,k_0}(z_i-z_j)$. Their explicit formulas and the operator-norm remainder estimates are given in Appendix~\ref{app:kernel-remainders}. The leading logarithmic and Newtonian self-interactions are local; the Bloch dependence enters through the regular quasiperiodic part and the interactions between distinct inclusions.

For subsequent use, define the block-diagonal operators on $\mathcal H$ by
\begin{equation} \label{eq:leading-block-operators}
    \mathcal P_{\mathrm{diag}}:=\bigoplus_{i=1}^N\mathcal P_i\quad(d=2),
    \qquad
    \mathcal N:=\bigoplus_{i=1}^N\mathcal N_i\quad(d=3).
\end{equation}
Thus $(\mathcal P_{\mathrm{diag}}U)_i=\mathcal P_iU_i$ and $(\mathcal N U)_i=\mathcal N_iU_i$. The notation $\mathcal P_{\mathrm{diag}}$ distinguishes the leading two-dimensional operator from the inclusion set $\mathcal D$, the orthogonal projection $P$ and the finite-dimensional matrix $\mathsf D$ introduced below.

\begin{remark}[Terminology]
\label{rem:resonance-terminology}
    For fixed $\kappa$, we refer to the characteristic values of the quasiperiodic operator pencil $\mathcal{A}_\kappa(\cdot,\delta)$ as resonant Bloch frequencies. In the undamped case, these characteristic values are with real-valued Bloch eigenfrequencies. When damping is present, the pencil is non-self-adjoint and the resonant frequencies generally move into the complex plane. The term resonance is used throughout in this operator-pencil sense and should not be confused with scattering resonances of a finite open system.
\end{remark}

\begin{remark}
    The key to the asymptotic regime studied here is that the singular factor $\tau(\omega,\delta)=\delta^2\omega^2\xi(\omega)$ multiplies an operator of magnitude $\log\delta$ in two dimensions and of size one in three dimensions. Near a material pole, this balance produces characteristic values close to the pole, while the regular quasiperiodic part determines their splitting.
\end{remark}

\section{Subwavelength resonance asymptotics}
\label{sec:subwavelength-asymptotics}

We now reduce the operator pencil \eqref{eq:operator-pencil-corrected} to the modes selected by the leading self-interaction on each inclusion. The selected modes are dimension dependent. In two dimensions the large logarithmic part has rank one on each component and selects the normalized constant mode. In three dimensions the leading self-interaction is the Newtonian potential, so the selected modes are its isolated eigenfunctions rather than constants in general.

Throughout this section, $\kappa$ ranges over a compact set $K\Subset Y^*$ and $\omega$ ranges over a compact neighbourhood $\mathcal U$ on which the material law is meromorphic. We assume that this parameter set is uniformly separated from diffraction thresholds: for some $c_{\mathscr D}>0$,
\begin{equation} \label{eq:uniform-diffraction-separation}
    \inf_{\substack{\kappa\in K,\,\omega\in\mathcal U\\q\in\Lambda^*}}
    \left|k_0(\omega)^2-|\kappa+q|^2\right|
    \geq c_{\mathscr D}.
\end{equation}
The geometric operator $\mathcal K_\kappa(\omega,\delta)$ remains defined at material poles: its kernel depends on the background wavenumber, not on $\xi$. Its estimates are uniform on $K\times\mathcal U$. The pencil $\mathcal A_\kappa$ and the contrast-dependent Schur complement are used only away from material poles, and only where the stated complementary inverse exists.

\subsection{Selected modal subspace}

For $d=2$, define
\begin{equation} \label{eq:normalized-indicators}
    \varphi_i:=\chi_i:=\frac{\mathbf 1_{D_i}}{\sqrt{|D_i|}},
    \qquad i=1,\dots,N.
\end{equation}
For $d=3$, let $\nu_i\neq0$ be a simple isolated eigenvalue of the Newtonian potential $\mathcal N_i$ in \eqref{eq:correct-3d-self}, and let $\varphi_i$ be an associated normalized eigenfunction satisfying $\mathcal N_i\varphi_i=\nu_i\varphi_i$ and $\|\varphi_i\|_{L^2(D_i)}=1$. The existence and spectral properties of these Newtonian modes are classical; see~\cite{anderson1992spectral}. We embed each $\varphi_i$ in
$\mathcal H=\bigoplus_iL^2(D_i)$ by setting all other components equal to zero, and define
\begin{equation}
    E:=\operatorname{span}\{\varphi_1,\dots,\varphi_N\},
    \qquad
    P:\mathcal H\to E,
    \qquad
    Q:=I_{\mathcal H}-P,
\end{equation}
where $P$ is the orthogonal projection and $I_{\mathcal H}$ is the identity on $\mathcal H$.

\begin{lemma}[Dominant resonant modes]
\label{thm:dominant-resonant-modes}
Let $\tau(\omega,\delta)=\delta^2\omega^2\xi(\omega)$ and suppose that
\begin{equation} \label{eq:complement-invertibility}
    Q\mathcal A_\kappa(\omega,\delta)Q
    =I_{Q\mathcal H}-\tau(\omega,\delta)
    Q\mathcal K_\kappa(\omega,\delta)Q
\end{equation}
is invertible with uniformly bounded inverse in the spectral window under consideration. If $U_\kappa$ is a resonant field, then
\begin{equation} \label{eq:complement-field-formula}
    QU_\kappa
    =
    \tau(\omega,\delta)
    \big(Q\mathcal A_\kappa Q\big)^{-1}
    Q\mathcal K_\kappa P U_\kappa.
\end{equation}
Consequently,
\begin{equation} \label{eq:correct-modal-decomposition}
    U_\kappa
    =
    \sum_{i=1}^N q_i\varphi_i+\mathrm{Rem}_{\kappa,\delta}.
\end{equation}
In the two-dimensional logarithmic regime $\tau=O(|\log\delta|^{-1})$, one has
$\|\mathrm{Rem}_{\kappa,\delta}\|=O(|\log\delta|^{-1})\|PU_\kappa\|$. In three dimensions, for bounded $\tau$, one has
$\|\mathrm{Rem}_{\kappa,\delta}\|=O(\delta)\|PU_\kappa\|$.
\end{lemma}

\begin{proof}
Write $U_\kappa=PU_\kappa+QU_\kappa$. Applying $Q$ to $\mathcal A_\kappa U_\kappa=0$ and using $QP=0$ gives
\begin{equation}
    Q\mathcal A_\kappa P\,PU_\kappa
    +Q\mathcal A_\kappa Q\,QU_\kappa=0.
\end{equation}
Since $\mathcal A_\kappa=I-\tau\mathcal K_\kappa$,
\begin{equation}
    Q\mathcal A_\kappa P
    =QP-\tau Q\mathcal K_\kappa P
    =-\tau Q\mathcal K_\kappa P,
\end{equation}
and hence
\begin{equation}
    Q\mathcal A_\kappa Q\,QU_\kappa
    =\tau Q\mathcal K_\kappa P\,PU_\kappa.
\end{equation}
Invertibility of $Q\mathcal A_\kappa Q$ now yields
\begin{equation}
    QU_\kappa
    =\tau(Q\mathcal A_\kappa Q)^{-1}
    Q\mathcal K_\kappa P\,PU_\kappa,
\end{equation}
which is \eqref{eq:complement-field-formula}.

In two dimensions, \eqref{eq:correct-2d-self}--\eqref{eq:correct-2d-cross} imply
\begin{equation}
    \mathcal K_\kappa P
    =-\frac{\log\delta}{2\pi}\mathcal P_{\mathrm{diag}}P+\mathcal R_{\kappa,\delta}P,
    \qquad
    \|\mathcal R_{\kappa,\delta}P\|=O(1),
\end{equation}
where $\mathcal P_{\mathrm{diag}}P$ takes values in $E$. Therefore
\begin{equation}
    Q\mathcal K_\kappa P
    =Q\mathcal R_{\kappa,\delta}P,
    \qquad
    \|Q\mathcal K_\kappa P\|=O(1).
\end{equation}
Using the uniform inverse bound and $\tau=O(|\log\delta|^{-1})$ gives
\begin{equation}
    \|QU_\kappa\|
    \leq C|\tau|\,\|Q\mathcal K_\kappa P\|\,\|PU_\kappa\|
    =O(|\log\delta|^{-1})\|PU_\kappa\|.
\end{equation}

In three dimensions,
\begin{equation}
    \mathcal K_\kappa P
    =\mathcal N P+\delta\mathcal R_{\kappa,\delta}P,
    \qquad
    \mathcal N\varphi_i=\nu_i\varphi_i,
\end{equation}
so $Q\mathcal NP=0$ and
\begin{equation}
    \|Q\mathcal K_\kappa P\|=O(\delta).
\end{equation}
For bounded $\tau$ this gives
\begin{equation}
    \|QU_\kappa\|=O(\delta)\|PU_\kappa\|.
\end{equation}
Finally, $PU_\kappa=\sum_{i=1}^Nq_i\varphi_i$ and
$\mathrm{Rem}_{\kappa,\delta}:=QU_\kappa$, proving the decomposition and both estimates.
This is the explicit $P/Q$ form of the modal reduction used in the subwavelength volume-potential arguments of~\cite{ammari2022wave,ammari2024functional,AD,alexopoulos2022mathematical}.
\end{proof}

\subsection{Schur-complement reduction}

The complement-invertibility hypothesis excludes resonances associated with modes outside $E$ from the selected spectral window. It is the precise spectral-isolation assumption required for a one-mode-per-inclusion reduction.

\begin{lemma}[Finite-dimensional subwavelength reduction]
\label{prop:finite-dimensional-reduction}
Under the assumptions of Lemma~\ref{thm:dominant-resonant-modes}, define the Schur complement
\begin{equation} \label{eq:schur-complement}
    \mathcal S_\kappa
    :=
    P\mathcal A_\kappa P
    -
    P\mathcal A_\kappa Q
    \big(Q\mathcal A_\kappa Q\big)^{-1}
    Q\mathcal A_\kappa P.
\end{equation}
Then
\begin{equation}
    \ker\mathcal A_\kappa(\omega,\delta)\neq\{0\}
    \quad\Longleftrightarrow\quad
    \ker\mathcal S_\kappa(\omega,\delta)\neq\{0\}.
\end{equation}
In the basis $\{\varphi_i\}_{i=1}^N$,
\begin{equation} \label{eq:finite-resonance-system}
    \mathcal S_\kappa
    =I_N-\tau(\omega,\delta)
    \mathsf M_\kappa^{\mathrm{eff}}(\omega,\delta),
\end{equation}
where
\begin{equation} \label{eq:effective-matrix}
    \mathsf M_\kappa^{\mathrm{eff}}
    :=
    P\mathcal K_\kappa P
    +
    \tau P\mathcal K_\kappa Q
    \big(Q\mathcal A_\kappa Q\big)^{-1}
    Q\mathcal K_\kappa P.
\end{equation}
Thus the characteristic values in the selected spectral window are determined exactly by
\begin{equation} \label{eq:finite-det-condition}
    \det\left(
    I_N-\delta^2\omega^2\xi(\omega)
    \mathsf M_\kappa^{\mathrm{eff}}(\omega,\delta)
    \right)=0.
\end{equation}
Equivalently, if $\widehat\lambda_j^\kappa(\delta,\omega)$ is a simple eigenvalue of $\mathsf M_\kappa^{\mathrm{eff}}$, the associated branch satisfies
\begin{equation} \label{eq:eigenvalue-resonance-condition}
    \delta^2\omega^2\xi(\omega)
    \widehat\lambda_j^\kappa(\delta,\omega)=1.
\end{equation}
\end{lemma}

\begin{proof}
Relative to $\mathcal H=E\oplus Q\mathcal H$, the equation
$\mathcal A_\kappa U_\kappa=0$ is
\begin{equation}
    \begin{pmatrix}
        P\mathcal A_\kappa P & P\mathcal A_\kappa Q\\
        Q\mathcal A_\kappa P & Q\mathcal A_\kappa Q
    \end{pmatrix}
    \begin{pmatrix}PU_\kappa\\QU_\kappa\end{pmatrix}
    =\begin{pmatrix}0\\0\end{pmatrix}.
\end{equation}
The second block equation gives
\begin{equation} \label{eq:schur-complement-field-elimination}
    QU_\kappa
    =-(Q\mathcal A_\kappa Q)^{-1}
    Q\mathcal A_\kappa P\,PU_\kappa.
\end{equation}
Substitution into the first block equation yields
\begin{equation}
    \left[
    P\mathcal A_\kappa P
    -P\mathcal A_\kappa Q
    (Q\mathcal A_\kappa Q)^{-1}
    Q\mathcal A_\kappa P
    \right]PU_\kappa=0,
\end{equation}
that is, $\mathcal S_\kappa PU_\kappa=0$. Conversely, if
$0\neq V\in\ker\mathcal S_\kappa$, then
\begin{equation}
    U:=V-(Q\mathcal A_\kappa Q)^{-1}Q\mathcal A_\kappa P V
\end{equation}
satisfies $\mathcal A_\kappa U=0$. This proves the equivalence of the kernels.

Using the fact that
\begin{align}
    P\mathcal A_\kappa P=P-\tau P\mathcal K_\kappa P,\qquad
    P\mathcal A_\kappa Q=-\tau P\mathcal K_\kappa Q,\qquad
    Q\mathcal A_\kappa P=-\tau Q\mathcal K_\kappa P,
\end{align}
we have that
\begin{align}
    \mathcal S_\kappa
    &=P-\tau P\mathcal K_\kappa P
    -\tau^2P\mathcal K_\kappa Q
    (Q\mathcal A_\kappa Q)^{-1}
    Q\mathcal K_\kappa P\notag\\
    &=P-\tau\left[
    P\mathcal K_\kappa P
    +\tau P\mathcal K_\kappa Q
    (Q\mathcal A_\kappa Q)^{-1}
    Q\mathcal K_\kappa P
    \right].
\end{align}
After identifying $P|_E$ with $I_N$, this is precisely
\eqref{eq:finite-resonance-system}--\eqref{eq:effective-matrix}. Since $E$ has dimension $N$,
\begin{equation}
    \ker\mathcal S_\kappa\neq\{0\}
    \quad\Longleftrightarrow\quad
    \det\mathcal S_\kappa=0,
\end{equation}
which gives \eqref{eq:finite-det-condition}. Diagonalizing a simple eigenvalue
$\widehat\lambda_j^\kappa$ of $\mathsf M_\kappa^{\mathrm{eff}}$ then gives
$1-\tau\widehat\lambda_j^\kappa=0$, namely \eqref{eq:eigenvalue-resonance-condition}.
\end{proof}

\subsection{Dimension-dependent interaction matrices}

The Schur-complement matrix has different leading forms in two and three dimensions. These formulas make explicit where the Bloch dependence enters and replace a term-by-term expansion of the conditionally convergent real-space lattice sum.

\begin{theorem}[Two-dimensional interaction asymptotics]
\label{thm:two-dimensional-asymptotics}
Let $d=2$ and suppose that $\tau=O(|\log\delta|^{-1})$. Then
\begin{equation} \label{eq:effective-matrix-2d}
    \mathsf M_\kappa^{\mathrm{eff}}(\omega,\delta)
    =
    -\frac{\log\delta}{2\pi}\mathsf D
    +\mathsf B_\kappa^{(2)}(\omega)
    +O(|\log\delta|^{-1})+O(\delta),
\end{equation}
where $\mathsf D:=\operatorname{diag}(|D_1|,\dots,|D_N|)$ and
\begin{equation} \label{eq:2d-regular-interaction-matrix}
    \big(\mathsf B_\kappa^{(2)}(\omega)\big)_{ij}
    :=
    \left\langle
    \mathcal K_{ij}^{\kappa,\omega,(0)}\chi_j,
    \chi_i
    \right\rangle_{L^2(D_i)}.
\end{equation}
In particular, for $i\neq j$,
\begin{equation}
    \big(\mathsf B_\kappa^{(2)}(\omega)\big)_{ij}
    =
    -G^{\kappa,k_0}(z_i-z_j)
    \sqrt{|D_i||D_j|}.
\end{equation}
The leading logarithmic displacement is local and independent of $\kappa$; the regular matrix $\mathsf B_\kappa^{(2)}$ produces the splitting.
\end{theorem}

\begin{proof}
The matrix of $P\mathcal K_\kappa P$ in the orthonormal basis
$\{\chi_i\}_{i=1}^N$ has entries
\begin{equation}
    (P\mathcal K_\kappa P)_{ij}
    =\left\langle
    \mathcal K_{ij}^{\kappa,\omega,\delta}\chi_j,
    \chi_i
    \right\rangle_{L^2(D_i)}.
\end{equation}
For $i=j$, equation \eqref{eq:correct-2d-self} gives
\begin{align}
    (P\mathcal K_\kappa P)_{ii}
    &=-\frac{\log\delta}{2\pi}\langle\mathcal P_i\chi_i,\chi_i\rangle
    +\left\langle
    \mathcal K_{ii}^{\kappa,\omega,(0)}\chi_i,
    \chi_i
    \right\rangle+O(\delta).
\end{align}
Since $\int_{D_i}\chi_i=\sqrt{|D_i|}$ and $mathcal P_i\chi_i=\sqrt{|D_i|}\,\mathbf1_{D_i}$ one obtains $\langle\mathcal P_i\chi_i,\chi_i\rangle=|D_i|$. For $i\neq j$, equation \eqref{eq:correct-2d-cross} yields
\begin{align}
    (P\mathcal K_\kappa P)_{ij}
    &=\left\langle
    \mathcal K_{ij}^{\kappa,\omega,(0)}\chi_j,
    \chi_i
    \right\rangle+O(\delta)\notag\\
    &=-G^{\kappa,k_0}(z_i-z_j)
    \left(\int_{D_j}\chi_j\right)
    \overline{\left(\int_{D_i}\chi_i\right)}+O(\delta)\notag\\
    &=-G^{\kappa,k_0}(z_i-z_j)
    \sqrt{|D_i||D_j|}+O(\delta).
\end{align}
Therefore
\begin{equation}
    P\mathcal K_\kappa P
    =-\frac{\log\delta}{2\pi}\mathsf D+\mathsf B_\kappa^{(2)}+O(\delta).
\end{equation}

It remains to estimate the Schur correction. From the proof of
Lemma~\ref{thm:dominant-resonant-modes} we have that $\|Q\mathcal K_\kappa P\|+\|P\mathcal K_\kappa Q\|=O(1)$. Together with the estimates $\|(Q\mathcal A_\kappa Q)^{-1}\|\leq C$ and
$\tau=O(|\log\delta|^{-1})$, this gives
\begin{align}
    \left\|
    \tau P\mathcal K_\kappa Q
    (Q\mathcal A_\kappa Q)^{-1}
    Q\mathcal K_\kappa P
    \right\|\leq
    |\tau|\,\|P\mathcal K_\kappa Q\|
    \|(Q\mathcal A_\kappa Q)^{-1}\|
    \|Q\mathcal K_\kappa P\|
    =O(|\log\delta|^{-1}).
\end{align}
Substitution into \eqref{eq:effective-matrix} proves
\eqref{eq:effective-matrix-2d}.
\end{proof}

\begin{theorem}[Three-dimensional interaction asymptotics]
\label{thm:three-dimensional-asymptotics}
Let $d=3$ and suppose that
\begin{align*}
|\tau(\omega,\delta)|\leq C_\tau
\qquad \text{ and } \qquad
\bigl\|(Q\mathcal A_\kappa Q)^{-1}\bigr\|\leq C_Q,
\end{align*}
where the constants are independent of $\omega$, $\delta$
and the Bloch parameter $\kappa$. Then
\begin{equation} \label{eq:effective-matrix-3d}
    \mathsf M_\kappa^{\mathrm{eff}}(\omega,\delta)
    =
    \operatorname{diag}(\nu_1,\dots,\nu_N)
    +\delta\mathsf B_\kappa^{(3)}(\omega)
    +O(\delta^2),
\end{equation}
where
\begin{equation}
    \big(\mathsf B_\kappa^{(3)}(\omega)\big)_{ij}
    :=
    \left\langle
    \mathcal K_{ij}^{\kappa,\omega,(1)}\varphi_j,
    \varphi_i
    \right\rangle_{L^2(D_i)}.
\end{equation}
For $i\neq j$ this gives
\begin{equation}
    \big(\mathsf B_\kappa^{(3)}(\omega)\big)_{ij}
    =
    -G^{\kappa,k_0}(z_i-z_j)
    \bigg(\int_{D_j}\varphi_j\bigg)
    \overline{\bigg(\int_{D_i}\varphi_i\bigg)}.
\end{equation}
\end{theorem}

\begin{proof}
As in the two-dimensional case,
\begin{equation}
    (P\mathcal K_\kappa P)_{ij}
    =\left\langle
    \mathcal K_{ij}^{\kappa,\omega,\delta}\varphi_j,
    \varphi_i
    \right\rangle_{L^2(D_i)}.
\end{equation}
For $i=j$, equation \eqref{eq:correct-3d-self} and
$\mathcal N_i\varphi_i=\nu_i\varphi_i$ give
\begin{align}
    (P\mathcal K_\kappa P)_{ii}
    &=\langle\mathcal N_i\varphi_i,\varphi_i\rangle
    +\delta\left\langle
    \mathcal K_{ii}^{\kappa,\omega,(1)}\varphi_i,
    \varphi_i
    \right\rangle+O(\delta^2)\notag\\
    &=\nu_i
    +\delta\big(\mathsf B_\kappa^{(3)}\big)_{ii}
    +O(\delta^2).
\end{align}
For $i\neq j$, equation \eqref{eq:correct-3d-cross} gives
\begin{align}
    (P\mathcal K_\kappa P)_{ij}
    &=\delta\left\langle
    \mathcal K_{ij}^{\kappa,\omega,(1)}\varphi_j,
    \varphi_i
    \right\rangle+O(\delta^2)\notag\\
    &=-\delta G^{\kappa,k_0}(z_i-z_j)
    \left(\int_{D_j}\varphi_j\right)
    \overline{\left(\int_{D_i}\varphi_i\right)}
    +O(\delta^2).
\end{align}
Consequently,
\begin{equation}
    P\mathcal K_\kappa P
    =\operatorname{diag}(\nu_1,\dots,\nu_N)
    +\delta\mathsf B_\kappa^{(3)}+O(\delta^2).
\end{equation}

Moreover, it holds that $\|Q\mathcal K_\kappa P\|+\|P\mathcal K_\kappa Q\|=O(\delta)$. Since $\tau$ and $(Q\mathcal A_\kappa Q)^{-1}$ are uniformly bounded,
\begin{align}
    &\left\|
    \tau P\mathcal K_\kappa Q
    (Q\mathcal A_\kappa Q)^{-1}
    Q\mathcal K_\kappa P
    \right\|=O(\delta^2).
\end{align}
Inserting these two estimates into \eqref{eq:effective-matrix} proves
\eqref{eq:effective-matrix-3d}. The spectral properties of the Newtonian potential used here are described in~\cite{anderson1992spectral}.
\end{proof}

\begin{remark} \label{rem:uniform-tau-near-pole}
    Uniform boundedness of $\tau=\delta^2\omega^2\xi(\omega)$ does not require a fixed positive separation from a material pole. Indeed, near a simple nonzero pole $\omega^*$, we can write
    \begin{align*}
        \xi(\omega)=\frac{r_*}{\omega-\omega^*}+O(1).
    \end{align*}
    For $\omega=\omega^*+c\delta^2$, with fixed $c\ne0$, we obtain
    \begin{align*}
        \tau(\omega,\delta) = \frac{(\omega^*)^2r_*}{c}+O(\delta^2).
    \end{align*}
    Thus, the boundedness assumption permits frequencies approaching the pole on the three-dimensional resonant scale. Hence, the pole itself is excluded and the uniform bound on the complementary inverse remains a separate assumption.
\end{remark}

\subsection{Analytic interaction eigenvalues}

The matrix $\mathsf M_\kappa^{\mathrm{eff}}$ is an exact reduction where the complementary inverse exists, but it contains the singular factor $\tau$. We do not evaluate this matrix at a material pole. Instead, we use eigenvalues $\lambda_j^\kappa(\delta,\omega)$ of the full geometric operator $\mathcal K_\kappa(\omega,\delta)$; the hat in $\widehat\lambda_j^\kappa$ distinguishes the contrast-dependent matrix eigenvalues above.

\begin{proposition}[Analytic geometric spectrum]
\label{prop:analytic-geometric-spectrum}
Under \eqref{eq:uniform-diffraction-separation}, for each fixed $\delta>0$ the family $\mathcal K_\kappa(\omega,\delta)$ is compact and holomorphic in $\omega$ throughout the interior of $\mathcal U$, including material poles. A simple non-zero eigenvalue has a local holomorphic continuation. If a selected continuation exists on a common neighbourhood and its rank-one Riesz projection $\Pi_j^\kappa(\delta,\omega)$ satisfies
\begin{equation} \label{eq:uniform-riesz-bound}
    \sup_{\kappa,\omega,\delta}\|\Pi_j^\kappa(\delta,\omega)\|<\infty,
\end{equation}
then, on a smaller compact neighbourhood,
\begin{equation} \label{eq:geometric-eigenvalue-derivative}
    |\partial_\omega\lambda_j^\kappa(\delta,\omega)|\leq C.
\end{equation}
Away from the material poles, the characteristic equation on this eigenvalue branch is exactly
\begin{equation} \label{eq:branch-equation}
    \delta^2\omega^2\xi(\omega)\lambda_j^\kappa(\delta,\omega)=1.
\end{equation}
\end{proposition}

\begin{proof}
The local logarithmic and Newtonian kernels are square integrable on $D_i\times D_i$ in dimensions two and three, respectively. All off-diagonal kernels are regular because the physical centres are separated. Thus $\mathcal K_\kappa$ is Hilbert--Schmidt. Away from the background spectrum, the quasiperiodic background resolvent is holomorphic in $k_0^2=\varepsilon_0\mu_0\omega^2$. Restriction to the inclusions and rescaling preserve this holomorphy.

For a positively oriented contour $\Gamma_j$ enclosing only the selected eigenvalue, its spectral projection is
\begin{equation}
    \Pi_j^\kappa=\frac{1}{2\pi i}\int_{\Gamma_j}
    (zI-\mathcal K_\kappa)^{-1}\,\mathrm dz.
\end{equation}
Locally this projection is holomorphic and has rank one, and
\begin{equation}
    \lambda_j^\kappa=\operatorname{tr}(\mathcal K_\kappa\Pi_j^\kappa),
    \qquad
    \partial_\omega\lambda_j^\kappa
    =\operatorname{tr}((\partial_\omega\mathcal K_\kappa)\Pi_j^\kappa).
\end{equation}
To bound the derivative, write the frequency-independent leading terms as
\begin{equation}
    \mathcal K_\kappa(\omega,\delta)=
    \begin{cases}
    -\frac{\log\delta}{2\pi}\mathcal P_{\mathrm{diag}}+\mathcal R_{\kappa,\delta}(\omega),&d=2,\\
    \mathcal N+\delta\mathcal R_{\kappa,\delta}(\omega),&d=3,
    \end{cases}
    \qquad \|\mathcal R_{\kappa,\delta}(\omega)\|\leq C,
\end{equation}
with $\mathcal P_{\mathrm{diag}}$ and $\mathcal N$ defined in \eqref{eq:leading-block-operators}. The kernel estimates above hold uniformly on a slightly larger threshold-free complex neighbourhood. On circles of a fixed radius $r>0$ within that neighbourhood, Cauchy's formula gives
\begin{equation}
    \partial_\omega\mathcal R_{\kappa,\delta}(\omega)
    =\frac{1}{2\pi i}\int_{|s-\omega|=r}
    \frac{\mathcal R_{\kappa,\delta}(s)}{(s-\omega)^2}\,\mathrm ds,
    \qquad
    \|\partial_\omega\mathcal R_{\kappa,\delta}(\omega)\|\leq C/r.
\end{equation}
Thus $\|\partial_\omega\mathcal K_\kappa\|=O(1)$ in two dimensions and $O(\delta)$ in three dimensions. The rank-one trace bound yields
\begin{equation}
    |\partial_\omega\lambda_j^\kappa|
    \leq\|\partial_\omega\mathcal K_\kappa\|
    \|\Pi_j^\kappa\|\leq C.
\end{equation}
Finally, for $\tau\neq0$,
\begin{equation}
    (I-\tau\mathcal K_\kappa)U=0
    \quad\Longleftrightarrow\quad
    \mathcal K_\kappa U=\tau^{-1}U.
\end{equation}
This proves \eqref{eq:branch-equation} without evaluating the contrast-dependent Schur complement at a pole.
\end{proof}

The link to the finite-dimensional approximation can also be made without $\xi$. For a geometric spectral parameter $z\notin\sigma(Q\mathcal K_\kappa Q)$, define
\begin{equation} \label{eq:geometric-schur-matrix}
    \mathsf T_\kappa(z,\omega,\delta)
    :=P\mathcal K_\kappa P+
    P\mathcal K_\kappa Q(zI_Q-Q\mathcal K_\kappa Q)^{-1}
    Q\mathcal K_\kappa P.
\end{equation}
Indeed, $\mathcal K_\kappa(p+q)=z(p+q)$, with $p\in E$ and $q\in Q\mathcal H$, gives
\begin{equation}
    q=(zI_Q-Q\mathcal K_\kappa Q)^{-1}Q\mathcal K_\kappa Pp,
    \qquad
    (\mathsf T_\kappa(z,\omega,\delta)-zI_N)p=0.
\end{equation}
Consequently, the geometric eigenvalues in this window are characterized exactly by
\begin{equation}
    \det(\mathsf T_\kappa(z,\omega,\delta)-zI_N)=0.
\end{equation}
At $z=\tau^{-1}$, where both expressions exist,
\begin{equation}
    (zI_Q-Q\mathcal K_\kappa Q)^{-1}
    =\tau(I_Q-\tau Q\mathcal K_\kappa Q)^{-1},
    \qquad
    \mathsf T_\kappa(\tau^{-1},\omega,\delta)
    =\mathsf M_\kappa^{\mathrm{eff}}(\omega,\delta).
\end{equation}
Thus the two reductions agree at a resonance, although their eigenvalues are not identified away from the resonance condition.

In two dimensions, $Q\mathcal K_\kappa Q=O(1)$, so for $|z|\asymp |\log\delta|$ the complementary resolvent is $O(|\log\delta|^{-1})$. In three dimensions it is $O(1)$ for $z$ uniformly separated from the unselected Newtonian spectrum. The preceding block estimates give, respectively,
\begin{align} \label{eq:geometric-schur-asymptotics}
    \mathsf T_\kappa(z,\omega,\delta)
    &=-\frac{\log\delta}{2\pi}\mathsf D+\mathsf B_\kappa^{(2)}(\omega)
    +O(|\log\delta|^{-1})+O(\delta), &&d=2,\\
    \mathsf T_\kappa(z,\omega,\delta) \label{eq:geometric-schur-asymptotics-3d}
    &=\operatorname{diag}(\nu_1,\dots,\nu_N)
    +\delta\mathsf B_\kappa^{(3)}(\omega)+O(\delta^2), &&d=3.
\end{align}
These estimates hold also at material poles. Simple eigenvalue continuation on a common neighbourhood, separation from the unselected spectrum, and the uniform projection bound remain explicit hypotheses; simplicity at a single parameter value alone does not imply their uniform versions.

\section{Bloch branches generated by a material pole}
\label{sec:material-singularities}
\label{sec:band-formation-gap-mechanisms}

We now use the exact geometric eigenvalue condition \eqref{eq:branch-equation} to determine how a real material pole organizes the Bloch spectrum. The displacement from the pole is governed by eigenvalues of the full quasiperiodic interaction operator, whose selected modes are approximated by the matrices in Section~\ref{sec:subwavelength-asymptotics}.

Throughout this section we consider the undamped case $\gamma=0$, so that the permittivity is real-valued for real frequencies away from its poles:
\begin{equation} \label{eq:undamped-permittivity}
    \varepsilon(\omega) = \varepsilon_0 + \frac{\alpha}{1-\beta\omega^2}.
\end{equation}
The corresponding contrast $\xi(\omega)$ and the material poles $\omega_\pm^{*}$ are given by
\begin{equation} \label{eq:undamped-contrast}
    \xi(\omega) = \frac{\mu_0\alpha}{1-\beta\omega^2} \qquad \text{and} \qquad \omega_\pm^* = \pm\frac{1}{\sqrt{\beta}},
\end{equation}
respectively.

We focus on one of these poles and denote it by $\omega^*$. Near $\omega^*$, the contrast has the Laurent expansion
\begin{equation} \label{eq:contrast-laurent}
    \xi(\omega)
    =
    \frac{r_*}{\omega-\omega^*}
    +
    \xi_{\mathrm{reg}}(\omega),
\end{equation}
where $\xi_{\mathrm{reg}}$ is analytic near $\omega^*$ and $ r_*:=\operatorname*{Res}_{\omega=\omega^*}\xi(\omega)$. For the Drude--Lorentz law \eqref{eq:undamped-contrast}, this residue is
\begin{equation}
    r_*
    =
    -\frac{\mu_0\alpha}{2\beta\omega^*}.
\end{equation}

\subsection{Resonance equation near a material pole}

For a selected eigenvalue branch of the full geometric operator, the resonant Bloch frequencies satisfy the exact equation
\begin{equation} \label{eq:singular-resonance-equation}
    \delta^2\omega^2\xi(\omega)\lambda_j^\kappa(\delta,\omega)
    =
    1,
\end{equation}
where $\lambda_j^\kappa(\delta,\omega)$ is an eigenvalue of $\mathcal K_\kappa(\omega,\delta)$, not of the contrast-dependent matrix $\mathsf M_\kappa^{\mathrm{eff}}$. Proposition~\ref{prop:analytic-geometric-spectrum} supplies its local analytic continuation through a material pole whenever the selected eigenvalue is simple.

Near a material pole, the dominant contribution comes from the singular part of $\xi$. Substituting \eqref{eq:contrast-laurent} into \eqref{eq:singular-resonance-equation} gives
\begin{equation}
    \delta^2\omega^2
    \left(
        \frac{r_*}{\omega-\omega^*}
        +
        \xi_{\mathrm{reg}}(\omega)
    \right)
    \lambda_j^\kappa(\delta,\omega)
    =
    1.
\end{equation}
Thus, to leading order,
\begin{equation}
    \omega-\omega^*
    \sim
    \delta^2\omega_*^2 r_*
    \lambda_j^\kappa(\delta,\omega^*).
\end{equation}
This shows that material poles generate resonant frequencies located close to $\omega^*$, with the displacement from the pole controlled by the interaction eigenvalues, which depend on the Bloch parameter.

\begin{theorem}[Resonant branches near a material pole]
\label{prop:branches-near-pole}
Let $\omega^*\neq0$ be a simple real pole of the undamped contrast $\xi$, with non-zero residue $r_*$, and let the compact set $K\Subset Y^*$ satisfy the uniform threshold-separation condition \eqref{eq:uniform-diffraction-separation} in a fixed complex neighbourhood $\mathcal U$ of $\omega^*$. Set
\begin{equation}
    \Lambda_\delta:=
    \begin{cases}
        -\frac{\log\delta}{2\pi}, & d=2,\\
        1, & d=3.
    \end{cases}
\end{equation}
Suppose that a simple non-zero eigenvalue branch $\lambda_j^\kappa(\delta,\omega)$ of $\mathcal K_\kappa(\omega,\delta)$ continues holomorphically throughout $\mathcal U$, continuously in $\kappa\in K$, with uniformly bounded rank-one Riesz projections as in \eqref{eq:uniform-riesz-bound}. Assume, for constants $c,C>0$ independent of $\delta$ and $\kappa$, that
\begin{equation} \label{eq:eigenvalue-uniform-pole-bounds}
    c\Lambda_\delta
    \leq
    \big|\lambda_j^\kappa(\delta,\omega^*)\big|
    \leq C\Lambda_\delta.
\end{equation}
Then, for $\delta>0$ sufficiently small, the scalar characteristic equation associated with this eigenvalue branch has a unique solution $\omega_j(\kappa)$ in a disc of radius $R\delta^2\Lambda_\delta$ about $\omega^*$, for some fixed sufficiently large $R$. This solution is a real Bloch eigenfrequency and satisfies
\begin{equation} \label{eq:branch-asymptotic-pole}
    \omega_j(\kappa)
    =
    \omega^*
    +
    \delta^2\omega_*^2 r_*
    \lambda_j^\kappa(\delta,\omega^*)
    +O\!\left(\delta^4\Lambda_\delta^2\right),
\end{equation}
uniformly for $\kappa\in K$.
\end{theorem}

\begin{proof}
By Proposition~\ref{prop:analytic-geometric-spectrum}, after shrinking $\mathcal U$ once if necessary,
\begin{equation} \label{eq:pole-eigenvalue-derivative-bound}
    \sup_{\kappa\in K,\,\omega\in\mathcal U}
    |\partial_\omega\lambda_j^\kappa(\delta,\omega)|
    \leq C\leq C'\Lambda_\delta
\end{equation}
for sufficiently small $\delta$. The exact resonance equation may therefore be written as
\begin{equation}
    F_j(\omega,\kappa,\delta)
    :=
    1
    -
    \delta^2\omega^2\xi(\omega)
    \lambda_j^\kappa(\delta,\omega)
    =
    0.
\end{equation}
Using the Laurent expansion \eqref{eq:contrast-laurent}, this becomes
\begin{equation}
    1
    -
    \delta^2\omega^2
    \left(
        \frac{r_*}{\omega-\omega^*}
        +
        \xi_{\mathrm{reg}}(\omega)
    \right)
    \lambda_j^\kappa(\delta,\omega)
    =
    0.
\end{equation}
Multiplying by $\omega-\omega^*$ gives the analytic equation
\begin{equation}
    H_j(\omega,\kappa,\delta):=
    \omega-\omega^*
    -
    \delta^2\omega^2r_*
    \lambda_j^\kappa(\delta,\omega)
    -
    \delta^2\omega^2(\omega-\omega^*)
    \xi_{\mathrm{reg}}(\omega)
    \lambda_j^\kappa(\delta,\omega)
    =
    0.
\end{equation}
Set
\begin{equation}
    \zeta:=\omega-\omega^*,
    \qquad
    \eta_\delta:=\delta^2\Lambda_\delta.
\end{equation}
Then $H_j=0$ is equivalent to the fixed-point equation
\begin{equation} \label{eq:undamped-pole-fixed-point}
    \zeta=\mathcal T_{\kappa,\delta}(\zeta),
\end{equation}
where
\begin{align}
    \mathcal T_{\kappa,\delta}(\zeta)
    :=\delta^2(\omega^*+\zeta)^2
    \lambda_j^\kappa(\delta,\omega^*+\zeta)
    \left[r_*+\zeta\xi_{\mathrm{reg}}(\omega^*+\zeta)\right].
\end{align}
Fix $R>0$ and let $B_{R\eta_\delta}:=\{\zeta:|\zeta|\leq
R\eta_\delta\}$. The bounds \eqref{eq:eigenvalue-uniform-pole-bounds} imply
\begin{equation}
    \sup_{\zeta\in B_{R\eta_\delta}}
    \left|\lambda_j^\kappa(\delta,\omega^*+\zeta)\right|
    \leq C_1\Lambda_\delta.
\end{equation}
Since $\xi_{\mathrm{reg}}$ is bounded near $\omega^*$, it follows that
\begin{equation}
    \sup_{\zeta\in B_{R\eta_\delta}}
    |\mathcal T_{\kappa,\delta}(\zeta)|
    \leq C_2\delta^2\Lambda_\delta
    =C_2\eta_\delta.
\end{equation}
Choosing $R>2C_2$ shows that $\mathcal T_{\kappa,\delta}$ maps
$B_{R\eta_\delta}$ into itself for small $\delta$.

Differentiating the fixed-point map gives
\begin{align}
    \mathcal T_{\kappa,\delta}'(\zeta)
    ={}&\delta^2\Big\{
    2(\omega^*+\zeta)\lambda_j^\kappa
    +(\omega^*+\zeta)^2\partial_\omega\lambda_j^\kappa
    \Big\}
    \left[r_*+\zeta\xi_{\mathrm{reg}}\right]+\delta^2(\omega^*+\zeta)^2\lambda_j^\kappa
    \left[\xi_{\mathrm{reg}}
    +\zeta\xi_{\mathrm{reg}}'\right],
\end{align}
where all functions on the right are evaluated at $\omega^*+\zeta$.
Using \eqref{eq:eigenvalue-uniform-pole-bounds}, \eqref{eq:pole-eigenvalue-derivative-bound} and
$|\zeta|=O(\eta_\delta)$ yields
\begin{equation}
    \sup_{\zeta\in B_{R\eta_\delta}}
    |\mathcal T_{\kappa,\delta}'(\zeta)|
    \leq C_3\delta^2\Lambda_\delta
    =C_3\eta_\delta<1
\end{equation}
for small $\delta$. Thus \eqref{eq:undamped-pole-fixed-point} has a unique solution in this disc.

Finally, subtract the leading value
$\delta^2\omega_*^2r_*\lambda_j^\kappa(\delta,\omega^*)$ from
\eqref{eq:undamped-pole-fixed-point}. Since
$|\zeta|=O(\eta_\delta)$, the mean-value estimates
\begin{align}
    |(\omega^*+\zeta)^2-\omega_*^2|&=O(\eta_\delta),\\
    \left|\lambda_j^\kappa(\delta,\omega^*+\zeta)
    -\lambda_j^\kappa(\delta,\omega^*)\right|
    &=O(\Lambda_\delta\eta_\delta),
\end{align}
give
\begin{align}
    \zeta
    -\delta^2\omega_*^2r_*
    \lambda_j^\kappa(\delta,\omega^*)
    =O(\eta_\delta^2).
\end{align}
Since $\eta_\delta^2=\delta^4\Lambda_\delta^2$, this is
\eqref{eq:branch-asymptotic-pole}. Moreover,
\begin{equation}
    H_j(\omega^*,\kappa,\delta)
    =-\delta^2\omega_*^2r_*\lambda_j^\kappa(\delta,\omega^*)\neq0,
\end{equation}
so the solution is not the material pole and solves the original characteristic equation. For real $\kappa$, the geometric operator satisfies
$\mathcal K_\kappa(\overline\omega,\delta)
=\mathcal K_\kappa(\omega,\delta)^*$ and is self-adjoint for real $\omega$ away from background thresholds. The simple eigenvalue continuation therefore satisfies
$\lambda_j^\kappa(\delta,\overline\omega)
=\overline{\lambda_j^\kappa(\delta,\omega)}$ near the real pole. Since the undamped contrast has the same conjugation symmetry, the conjugate of the unique root is also a root in the disc. Uniqueness implies $\omega_j(\kappa)\in\mathbb R$. Continuous dependence of the uniform contraction on $\kappa$ gives a continuous Bloch branch.
\end{proof}

\begin{remark}[Degeneracies and diffraction thresholds]
\label{rem:degeneracies-diffraction-thresholds}
    The uniformity statement in Theorem~\ref{prop:branches-near-pole} concerns selected eigenvalues of the full geometric operator. Away from degeneracies means that the selected eigenvalue remains algebraically simple on a common frequency neighbourhood and that its spectral projection satisfies \eqref{eq:uniform-riesz-bound}. At each parameter value it is isolated, but a uniform lower bound on eigenvalue separation alone would not control spectral projections for a non-self-adjoint family. Conversely, selected eigenvalue separations may shrink with $\delta$ while the projections remain bounded. The hypotheses explicitly exclude exceptional points at which this control fails. Away from diffraction thresholds means that the background cell resolvent is uniformly invertible; in particular, one excludes points where
    \begin{equation}
        |\kappa+q|^2 = k_0^2
    \end{equation}
    for some reciprocal lattice vector $q$. Condition \eqref{eq:uniform-diffraction-separation} excludes these points uniformly as $\omega$ varies in the spectral neighbourhood. At a threshold the quasiperiodic Green function changes singular structure, and the operator expansions used above need not hold in the same form.
\end{remark}

\begin{remark}
    Theorem~\ref{prop:branches-near-pole} associates a resonant frequency with each selected geometric eigenvalue satisfying its uniform continuation hypotheses. The shift from $\omega^*$ is proportional to $\delta^2\lambda_j^\kappa(\delta,\omega^*)$, so these branches collapse onto the material pole as $\delta\to0$. The real pole therefore generates a family of resonant branches whose locations are determined by the quasiperiodic interaction between the inclusions.

    If all $N$ selected geometric eigenvalues satisfy these hypotheses, one obtains $N$ associated branches near the pole. Higher internal modes can also be considered, but their separation, size and projection bounds must be checked separately; the present theorem does not assert uniform control over an infinite family.
\end{remark}

\subsection{Two-dimensional form of the singular asymptotics}

In dimension $d=2$, the geometric reduction \eqref{eq:geometric-schur-asymptotics} gives, uniformly for $|z|\asymp |\log\delta|$,
\begin{equation}
    \mathsf T_\kappa(z,\omega,\delta)
    =-\frac{\log\delta}{2\pi}\mathsf D+\mathsf B_\kappa^{(2)}(\omega)
    +O(|\log\delta|^{-1})+O(\delta).
\end{equation}
Let $\lambda_j^\kappa(\delta,\omega)$ be a selected eigenvalue of the full operator, characterized by $\det(\mathsf T_\kappa(\lambda_j^\kappa,\omega,\delta)-\lambda_j^\kappa I_N)=0$. Under the hypotheses of Theorem~\ref{prop:branches-near-pole}, the corresponding resonant frequency satisfies
\begin{equation} \label{eq:2d-singular-branch}
    \omega_j(\kappa)
    =
    \omega^*
    +
    \delta^2\omega_*^2r_*
    \lambda_j^\kappa(\delta,\omega^*)
    +O(\delta^4|\log\delta|^2),
\end{equation}
so the common displacement from the pole is typically of order $\delta^2|\log\delta|$. The Bloch dependence does not arise from a formal sum of logarithmic constants over the lattice. It is contained in the regular matrix $\mathsf B_\kappa^{(2)}(\omega^*)$ through the properly defined quasiperiodic Green function.

The scale of the displacement should be distinguished from the scale of the splitting. If the inclusions have equal area $|D_i|=A$, then
\begin{equation} \label{eq:equal-area-eigenvalue-expansion}
    \lambda_j^\kappa(\delta,\omega^*)
    =-\frac{\log\delta}{2\pi} A+b_j^\kappa+O(|\log\delta|^{-1})+O(\delta),
\end{equation}
where $b_j^\kappa$ is the corresponding eigenvalue of $\mathsf B_\kappa^{(2)}(\omega^*)$. To see the eigenvalue estimate, take a unit vector $p$ in the kernel of $\mathsf T_\kappa(\lambda_j^\kappa,\omega^*,\delta)-\lambda_j^\kappa I_N$. Then
\begin{equation}
    \|[(\lambda_j^\kappa+\tfrac{\log\delta}{2\pi} A)I_N-\mathsf B_\kappa^{(2)}(\omega^*)]p\|
    \leq C(|\log\delta|^{-1}+\delta).
\end{equation}
For real $\omega^*$ and $\kappa$, the matrix $\mathsf B_\kappa^{(2)}(\omega^*)$ is Hermitian, so its spectral theorem bounds the distance of $\lambda_j^\kappa+\frac{\log\delta}{2\pi} A$ to its spectrum by the same quantity. For an isolated selected eigenvalue $b_j^\kappa$, this identifies the branch in \eqref{eq:equal-area-eigenvalue-expansion}. No evaluation of $\mathsf M_\kappa^{\mathrm{eff}}$ at $\omega^*$ is used. Hence
\begin{equation} \label{eq:equal-area-branch-expansion}
    \omega_j(\kappa)
    =\omega^*- \frac{\delta^2\log\delta}{2\pi}C_* A
    +\delta^2C_*b_j^\kappa+o(\delta^2),
    \qquad C_*:=\omega_*^2r_*.
\end{equation}
The logarithmic displacement is then common to all selected branches and cancels from their differences; their splitting is of order $\delta^2$. If the inclusion areas differ, the diagonal matrix $|\log\delta|\mathsf D$ can already separate the branches, and the corresponding splitting may instead be of order $\delta^2|\log\delta|$.

\subsection{A conditional complete-gap criterion}

Each resonant branch $\omega_j:Y^*\to\mathbb{R}$ corresponds to the Bloch band
\begin{equation} \label{eq:resonant-band-correspondence}
    B_j=\omega_j(Y^*)
    =\big\{\omega_j(\kappa):\kappa\in Y^*\big\}.
\end{equation}
Thus the branch asymptotics obtained above determine the location and width of the associated spectral bands. In particular, two neighbouring bands are separated by a complete gap whenever
\begin{equation} \label{eq:complete-gap-separation-concise}
    \sup_{\kappa\in Y^*}\omega_j(\kappa)
    <
    \inf_{\kappa\in Y^*}\omega_{j+1}(\kappa),
\end{equation}
provided that no other relevant band intersects the resulting interval.

The distinction between complete and partial gaps is relevant as, in numerical band diagrams, one often plots the dispersion relation only along a high-symmetry path in the Brillouin zone, for example $\Gamma \to X \to M \to \Gamma$ in a square lattice, as shown in the central panel of Figure~\ref{fig:constant-real-bands}. A gap visible along such a path means that no branch crosses the corresponding frequency interval for Bloch parameters on that path. However, another branch may cross the same interval at a Bloch parameter away from the plotted path.

The local pole asymptotics describe selected branches on threshold-free sets of Bloch parameters. A complete gap requires additional global information: separation of neighbouring bands over the entire Brillouin zone and exclusion of all other bands from the proposed interval. The following conditional criterion states the conclusion when this information is available independently.

\begin{proposition}[Conditional complete-gap criterion near a material pole]
\label{thm:gap-opening-near-pole}
Let $\omega_j(\kappa)$ and $\omega_{j+1}(\kappa)$ be two real, continuous Bloch branches in a fixed neighbourhood of a real material pole. Suppose that expansions over the full Brillouin zone have been established independently: for some $s_\delta>0$ with $s_\delta\to0$ and some common centre $\omega_{\mathrm c}(\delta)$,
\begin{equation} \label{eq:uniform-gap-branch-expansions}
    \omega_\ell(\kappa)
    =\omega_{\mathrm c}(\delta)
    +s_\delta a_\ell(\kappa)+r_{\ell,\delta}(\kappa),
    \qquad \ell=j,j+1,
\end{equation}
where $a_j,a_{j+1}$ are real and continuous on $Y^*$ and $\sup_{\kappa\in Y^*}|r_{\ell,\delta}(\kappa)|=o(s_\delta)$. Suppose that these are consecutive branches of the full Bloch spectrum in this neighbourhood and that
\begin{equation} \label{eq:eigenvalue-separation}
    \sup_{\kappa\in Y^*}a_j(\kappa)
    <
    \inf_{\kappa\in Y^*}a_{j+1}(\kappa).
\end{equation}
Then, for $\delta>0$ sufficiently small, a complete band gap opens between the two bands, and
\begin{equation} \label{eq:gap-width-asymptotic}
    \operatorname{gap}_{j,j+1}
    =s_\delta
    \left[
        \inf_{\kappa\in Y^*}a_{j+1}(\kappa)
        -
        \sup_{\kappa\in Y^*}a_j(\kappa)
    \right]
    +o(s_\delta).
\end{equation}
\end{proposition}

\begin{proof}
Introduce the notation $A_j^+:=\sup_{\kappa\in Y^*}a_j(\kappa)$, $A_{j+1}^-:=\inf_{\kappa\in Y^*}a_{j+1}(\kappa)$ and $R_{\ell,\delta}:=\sup_{\kappa\in Y^*}|r_{\ell,\delta}(\kappa)|=o(s_\delta)$. From \eqref{eq:uniform-gap-branch-expansions} we have that $ \sup_{\kappa\in Y^*}\omega_j(\kappa)\leq\omega_{\mathrm c}(\delta)+s_\delta A_j^++R_{j,\delta}$ and $\inf_{\kappa\in Y^*}\omega_{j+1}(\kappa)\geq\omega_{\mathrm c}(\delta)+s_\delta A_{j+1}^--R_{j+1,\delta}$. Let $\Delta A:=A_{j+1}^--A_j^+>0$, then taking the difference between these two inequalities leads to
\begin{align}
    \inf_{\kappa\in Y^*}\omega_{j+1}(\kappa)
    -\sup_{\kappa\in Y^*}\omega_j(\kappa)
    \geq s_\delta\Delta A
    -R_{j,\delta}-R_{j+1,\delta}
    =s_\delta\big(\Delta A-o(1)\big)>0
\end{align}
for sufficiently small $\delta$. Hence the two bands are disjoint. Because they are consecutive in the full Bloch spectrum, the open interval between these two edges is a complete gap. Moreover, the elementary inequalities
\begin{align}
    \left|
    \sup_{\kappa}(s_\delta a_j+r_{j,\delta})
    -s_\delta\sup_{\kappa}a_j
    \right|\leq R_{j,\delta}
    \qquad\text{and}\qquad
    \left|
    \inf_{\kappa}(s_\delta a_{j+1}+r_{j+1,\delta})
    -s_\delta\inf_{\kappa}a_{j+1}
    \right|\leq R_{j+1,\delta}
\end{align}
show that
\begin{align}
    \sup_{\kappa}\omega_j(\kappa)
    =\omega_{\mathrm c}(\delta)+s_\delta A_j^++o(s_\delta)
    \qquad\text{and}\qquad
    \inf_{\kappa}\omega_{j+1}(\kappa)
    =\omega_{\mathrm c}(\delta)+s_\delta A_{j+1}^-+o(s_\delta).
\end{align}
Taking the difference between these two expressions gives \eqref{eq:gap-width-asymptotic}.
\end{proof}

\begin{remark}
    The scale $s_\delta$ must resolve the first non-common term of the two branches. In three dimensions, distinct leading Newtonian eigenvalues give separation at order $\delta^2$. For equal leading Newtonian eigenvalues, the common order-$\delta^2$ displacement cancels and the first interaction splitting can instead occur at order $\delta^3$, according to \eqref{eq:geometric-schur-asymptotics} and its three-dimensional counterpart. In two dimensions, unequal leading inclusion areas may give $s_\delta=\delta^2|\log\delta|$. For equal-area inclusions, the common logarithmic displacement cancels by \eqref{eq:equal-area-branch-expansion}, and $s_\delta=\delta^2$ resolves the first regular splitting. These local scales yield a complete-gap conclusion only if the corresponding expansions and exclusion of other bands hold globally as assumed in the criterion.
\end{remark}

\begin{remark}
    Theorem~\ref{thm:gap-opening-near-pole} is a conditional criterion, not an existence theorem for complete gaps in the geometries studied here. Theorem~\ref{prop:branches-near-pole} is uniform only on selected compact sets away from background thresholds and degeneracies. At every positive real background wavenumber, some Bloch parameter satisfies $|\kappa+q|=k_0$: choose a vector of length $k_0$ and reduce it modulo the reciprocal lattice. One therefore cannot apply the threshold-free estimate unchanged over all of $Y^*$. A global gap proof would require additional analysis across those parameters and exclusion of any other branches from the candidate gap.
\end{remark}

This interpretation is the starting point for the damped problem. Damping moves the material pole and the associated resonant branches away from the real axis; the self-adjoint real-gap description then gives way to a complex spectrum and a frequency-dependent attenuation profile.

\section{Regularization of the singular Bloch spectrum by damping}
\label{sec:material-damping}

We now study how damping modifies the pole-induced branches described in Section~\ref{sec:material-singularities}. The material pole moves into the lower half-plane and the nearby resonant branches follow it. The real-axis singularity is thereby regularized, while the resonances persist as complex characteristic values. On the real-frequency slice, the resulting spectrum is described by directional attenuation rather than by a self-adjoint gap alone.

\subsection{Damped material law}

We return to the full Drude--Lorentz contrast
\begin{equation} \label{eq:damped-contrast}
    \xi(\omega)
    =
    \frac{\mu_0\alpha}{1-\beta\omega^2-i\gamma\omega},
    \qquad
    \gamma>0.
\end{equation}
The poles are
\begin{equation} \label{eq:damped-poles}
    \omega_\pm^*
    =
    \frac{1}{2\beta}
    \left(
        -i\gamma
        \pm
        \sqrt{4\beta-\gamma^2}
    \right).
\end{equation}
For small positive damping, these poles lie close to the real axis but have negative imaginary part:
\begin{equation}
    \operatorname{Im}\omega_\pm^*
    =
    -\frac{\gamma}{2\beta}
    <0.
\end{equation}
Thus, for real frequencies $\omega\in\mathbb{R}$, the contrast $\xi(\omega)$ remains finite. This is the fundamental difference from the undamped case.

Let
\begin{align*}
    \omega_0
    :=
    \operatorname{Re}\omega_+^*
    =
    \frac{1}{2\beta}
    \sqrt{4\beta-\gamma^2},
    \qquad
    \eta
    :=
    \frac{\gamma}{2\beta}.
\end{align*}
Then the pole can be written as $\omega_+^* = \omega_0 - i\eta$. Near $\omega_0$, the singular part of the contrast has the form
\begin{equation} \label{eq:damped-laurent-real-axis}
    \xi(\omega)
    =
    \frac{r_\gamma}{\omega-\omega_0+i\eta}
    +
    \xi_{\mathrm{reg},\gamma}(\omega),
\end{equation}
where $r_\gamma$ is the residue at the damped pole and $\xi_{\mathrm{reg},\gamma}$ is regular near $\omega_0$.

\subsection{Regularization of the resonance condition}

The subwavelength resonant branches are still determined by \eqref{eq:singular-resonance-equation}.
However, for real $\omega$, the factor $\xi(\omega)$ no longer blows up. Indeed, from \eqref{eq:damped-laurent-real-axis},
\begin{equation}
    \left|
        \frac{1}{\omega-\omega_0+i\eta}
    \right|
    =
    \frac{1}{\sqrt{(\omega-\omega_0)^2+\eta^2}},
\end{equation}
which remains bounded by $1/\eta$ on the real axis.

\begin{remark}
    Thus, damping replaces the singular blow-up $\frac{1}{\omega-\omega^*}$ by the regularized expression $\frac{1}{\omega-\omega_0+i\eta}$. The parameter $\eta$ measures the distance of the material pole from the real axis and controls the width of this regularization.
\end{remark}

\begin{theorem}[Damping regularizes the resonant Bloch branches]
\label{prop:damping-regularizes}
Let $\alpha,\beta,\mu_0>0$, $0<\gamma<2\sqrt{\beta}$, and let $\xi(\omega)$ be given by \eqref{eq:damped-contrast}. Denote by
\begin{equation}
    \omega_+^\gamma
    = \frac{1}{2\beta} \left( -i\gamma+\sqrt{4\beta-\gamma^2} \right)
\end{equation}
the damped pole bifurcating from the positive real pole
$\omega^*=1/\sqrt{\beta}$.

Let the compact subset $K\Subset Y^*$ be such that the condition \eqref{eq:uniform-diffraction-separation} holds in a neighbourhood of $\omega_+^\gamma$, and set
\begin{equation}
    \Lambda_\delta:=
    \begin{cases}
        -\frac{\log\delta}{2\pi}, & d=2,\\
        1, & d=3.
    \end{cases}
\end{equation}
Suppose that a simple non-zero eigenvalue branch $\lambda_j^\kappa(\delta,\omega)$ of $\mathcal K_\kappa(\omega,\delta)$ continues holomorphically on a common neighbourhood of $\omega_+^\gamma$, continuously in $\kappa\in K$, with uniformly bounded rank-one Riesz projections as in \eqref{eq:uniform-riesz-bound}. Assume that \eqref{eq:eigenvalue-uniform-pole-bounds} holds with $\omega^*$ replaced by $\omega_+^\gamma$. %Then, for $\delta>0$ sufficiently small, its scalar characteristic equation has a unique solution $\omega_j^\gamma(\kappa)$ in a disc of radius $R\delta^2\Lambda_\delta$ about $\omega_+^\gamma$, for some fixed sufficiently large $R$, satisfying
Then there exist constants $R>0$ and $\delta_0>0$, independent of $\delta$ and $\kappa$, such that, for every $0<\delta<\delta_0$ and $\kappa\in K$, the scalar characteristic equation associated with the selected eigenvalue branch has exactly one solution $\omega_j^\gamma(\kappa)$ in the disc $|\omega-\omega_+^\gamma|<R\delta^2\Lambda_\delta$. This solution satisfies
\begin{equation}
\label{eq:damped-branch-asymptotic}
    \omega_j^\gamma(\kappa)
    = \omega_+^\gamma + \delta^2 (\omega_+^\gamma)^2 r_\gamma \lambda_j^\kappa(\delta,\omega_+^\gamma) +O(\delta^4\Lambda_\delta^2),
\end{equation}
uniformly for $\kappa\in K$. Here
\begin{equation}
    r_\gamma
    = \operatorname*{Res}_{\omega=\omega_+^\gamma}\xi(\omega)
    = \frac{\mu_0\alpha}{-2\beta\omega_+^\gamma-i\gamma}.
\end{equation}
In particular,
\begin{equation}
\label{eq:damped-imaginary-part-scale}
    \operatorname{Im}\omega_j^\gamma(\kappa)
    = -\frac{\gamma}{2\beta}
    +
    \begin{cases}
        O(\delta^2|\log\delta|), & d=2,\\
        O(\delta^2), & d=3,
    \end{cases}
\end{equation}
and hence, for fixed $\gamma>0$ and sufficiently small $\delta$, the resonant Bloch branches lie strictly below the real frequency axis. Provided that the hypotheses above hold with constants uniform in $\gamma$, the same conclusion holds for $\gamma=\gamma(\delta)\to0$ if
\begin{equation} \label{eq:damping-dominates-interaction}
    \frac{\gamma(\delta)}{\delta^2\Lambda_\delta}\longrightarrow\infty.
\end{equation}
\end{theorem}

\begin{proof}
The poles of the damped contrast are the zeros of
\begin{equation}
    1-\beta\omega^2-i\gamma\omega.
\end{equation}
For $0<\gamma<2\sqrt{\beta}$, the pole bifurcating from the positive undamped pole $\omega^*=1/\sqrt{\beta}$ is
\begin{equation}
    \omega_+^\gamma
    = \frac{1}{2\beta} \left( -i\gamma+\sqrt{4\beta-\gamma^2} \right),
\end{equation}
so that
\begin{equation}
    \operatorname{Im}\omega_+^\gamma
    = -\frac{\gamma}{2\beta}<0.
\end{equation}
Thus, the material pole is displaced into the lower half-plane.

Near $\omega_+^\gamma$, the contrast has the Laurent expansion
\begin{equation}
    \xi(\omega)
    = \frac{r_\gamma}{\omega-\omega_+^\gamma} + \xi_{\mathrm{reg},\gamma}(\omega),
\end{equation}
where $\xi_{\mathrm{reg},\gamma}$ is analytic near $\omega_+^\gamma$ and
\begin{equation}
    r_\gamma
    = \operatorname*{Res}_{\omega=\omega_+^\gamma}\xi(\omega)
    = \frac{\mu_0\alpha} {-2\beta\omega_+^\gamma-i\gamma}.
\end{equation}

Proposition~\ref{prop:analytic-geometric-spectrum} supplies the derivative bound \eqref{eq:pole-eigenvalue-derivative-bound} on a smaller neighbourhood of $\omega_+^\gamma$. The damped resonant branches are determined exactly by
\begin{equation}
    \delta^2\omega^2\xi(\omega) \lambda_j^\kappa(\delta,\omega) = 1.
\end{equation}
Substituting the Laurent expansion gives
\begin{equation}
    1 - \delta^2\omega^2 \left( \frac{r_\gamma}{\omega-\omega_+^\gamma} + \xi_{\mathrm{reg},\gamma}(\omega) \right) \lambda_j^\kappa(\delta,\omega)
    = 0.
\end{equation}
Multiplying by $\omega-\omega_+^\gamma$, we obtain
\begin{align}
    0=\omega-\omega_+^\gamma
    -\delta^2\omega^2 r_\gamma
    \lambda_j^\kappa(\delta,\omega)\notag-\delta^2\omega^2(\omega-\omega_+^\gamma)
    \xi_{\mathrm{reg},\gamma}(\omega)
    \lambda_j^\kappa(\delta,\omega).
\end{align}
Set $\zeta:=\omega-\omega_+^\gamma$ and $\eta_\delta:=\delta^2\Lambda_\delta$. Then, the equation is equivalent to
\begin{equation} \label{eq:damped-pole-fixed-point}
    \zeta=\mathcal T_{\kappa,\delta}^\gamma(\zeta),
\end{equation}
where
\begin{align}
    \mathcal T_{\kappa,\delta}^\gamma(\zeta)
    :=\delta^2(\omega_+^\gamma+\zeta)^2
    \lambda_j^\kappa(\delta,\omega_+^\gamma+\zeta)\left[
    r_\gamma+\zeta\xi_{\mathrm{reg},\gamma}
    (\omega_+^\gamma+\zeta)
    \right].
\end{align}
On the disc $B_{R\eta_\delta}$, the uniform eigenvalue bounds and the boundedness of
$r_\gamma$, $\xi_{\mathrm{reg},\gamma}$ and their derivatives give
\begin{align}
    \sup_{B_{R\eta_\delta}}
    |\mathcal T_{\kappa,\delta}^\gamma|
    \leq C_1\delta^2\Lambda_\delta=C_1\eta_\delta,
    \qquad\text{and}\qquad
    \sup_{B_{R\eta_\delta}}
    |(\mathcal T_{\kappa,\delta}^\gamma)'|
    \leq C_2\delta^2\Lambda_\delta=C_2\eta_\delta.
\end{align}
Indeed, differentiating explicitly gives
\begin{align}
    (\mathcal T_{\kappa,\delta}^\gamma)'(\zeta)
    =\delta^2\Big(
    2(\omega_+^\gamma+\zeta)\lambda_j^\kappa
    +(\omega_+^\gamma+\zeta)^2\partial_\omega\lambda_j^\kappa\Big)\left[r_\gamma+\zeta\xi_{\mathrm{reg},\gamma}\right]+\delta^2(\omega_+^\gamma+\zeta)^2\lambda_j^\kappa
    \left[
    \xi_{\mathrm{reg},\gamma}
    +\zeta\xi_{\mathrm{reg},\gamma}'
    \right],
\end{align}
with all functions evaluated at $\omega_+^\gamma+\zeta$. %Choosing $R>2C_1$ and then $\delta$ small enough that $C_2\eta_\delta<1$ makes $\mathcal T_{\kappa,\delta}^\gamma$ a contraction of $B_{R\eta_\delta}$ into itself. Hence \eqref{eq:damped-pole-fixed-point} has a unique solution there. At $\zeta=0$ the right-hand side is $\delta^2(\omega_+^\gamma)^2r_\gamma\lambda_j^\kappa(\delta,\omega_+^\gamma)\neq0$, so this solution is distinct from the material pole and solves the original pencil equation. Uniform contraction and continuity in $\kappa$ give a continuous branch.
Let $\rho>0$ be such that the preceding estimates hold on $|\zeta|\leq\rho$, with constants $C_1,C_2$ independent of $\delta$ and $\kappa$. Fix $R>2C_1$ and choose $\delta_0>0$ sufficiently small that, for every $0<\delta<\delta_0$, $R\eta_\delta<\rho$ and $C_2\eta_\delta\leq\frac{1}{2}$. The map $\mathcal T_{\kappa,\delta}^\gamma$ then maps the closed disc $\overline{B_{R\eta_\delta}}$ into itself and is a contraction there. Therefore, the Banach fixed-point theorem gives exactly one solution of \eqref{eq:damped-pole-fixed-point} in this closed disc. Moreover, this solution satisfies
\begin{equation*}
    |\zeta|
    =|\mathcal T_{\kappa,\delta}^\gamma(\zeta)|
    \leq C_1\eta_\delta <R\eta_\delta,
\end{equation*}
so it lies in the open disc $B_{R\eta_\delta}$. Since
\begin{equation*}
    \mathcal T_{\kappa,\delta}^\gamma(0)
    =\delta^2(\omega_+^\gamma)^2r_\gamma \lambda_j^\kappa(\delta,\omega_+^\gamma)
    \neq0,
\end{equation*}
the solution is distinct from the material pole and satisfies the original scalar characteristic equation. The uniform contraction bound and continuity of $\mathcal T_{\kappa,\delta}^\gamma$ in $\kappa$ imply that the resulting resonant branch $\omega_j^\gamma(\kappa)$ is continuous in $\kappa$.

Subtracting the value of the fixed-point map at $\zeta=0$ gives
\begin{align}
    \zeta
    -\delta^2(\omega_+^\gamma)^2r_\gamma
    \lambda_j^\kappa(\delta,\omega_+^\gamma)
    =\mathcal T_{\kappa,\delta}^\gamma(\zeta)
    -\mathcal T_{\kappa,\delta}^\gamma(0)=\int_0^1
    (\mathcal T_{\kappa,\delta}^\gamma)'(t\zeta)\,\zeta\,\mathrm dt.
\end{align}
Therefore
\begin{equation}
    \left|
    \zeta
    -\delta^2(\omega_+^\gamma)^2r_\gamma
    \lambda_j^\kappa(\delta,\omega_+^\gamma)
    \right|
    \leq C\eta_\delta|\zeta|
    =O(\eta_\delta^2),
\end{equation}
which proves \eqref{eq:damped-branch-asymptotic}.

Taking imaginary parts in \eqref{eq:damped-branch-asymptotic} yields
\begin{equation}
    \operatorname{Im}\omega_j^\gamma(\kappa)
    = -\frac{\gamma}{2\beta}
    +\delta^2\operatorname{Im}\!\left[
        (\omega_+^\gamma)^2r_\gamma
        \lambda_j^\kappa(\delta,\omega_+^\gamma)
    \right]
    +O(\delta^4\Lambda_\delta^2),
\end{equation}
The eigenvalue bound gives
\begin{equation}
    \left|
    \delta^2\operatorname{Im}\!\left[
        (\omega_+^\gamma)^2r_\gamma
        \lambda_j^\kappa(\delta,\omega_+^\gamma)
    \right]
    \right|
    \leq C\delta^2\Lambda_\delta,
\end{equation}
while $\delta^4\Lambda_\delta^2=o(\delta^2\Lambda_\delta)$. Hence
\begin{equation}
    \operatorname{Im}\omega_j^\gamma(\kappa)
    =-\frac{\gamma}{2\beta}+O(\delta^2\Lambda_\delta),
\end{equation}
which is \eqref{eq:damped-imaginary-part-scale}. More explicitly, for some $C>0$,
\begin{equation}
    \operatorname{Im}\omega_j^\gamma(\kappa)
    \leq-\frac{\gamma}{2\beta}+C\delta^2\Lambda_\delta.
\end{equation}
For fixed $\gamma>0$, the right-hand side is negative for sufficiently small $\delta$. If $\gamma=\gamma(\delta)$ and
\eqref{eq:damping-dominates-interaction} holds, then
\begin{equation}
    \frac{C\delta^2\Lambda_\delta}{\gamma(\delta)/(2\beta)}
    \longrightarrow0,
\end{equation}
so the same strict negativity follows. Thus damping moves both the material pole and its associated resonant Bloch branches into the lower half-plane.
\end{proof}

\begin{remark}
    This theorem captures the main qualitative effect of damping: the material pole responsible for the singular branch organization is no longer situated on the real axis.
\end{remark}

\begin{remark}[Local complex-frequency and complex-Bloch descriptions]
    Theorem~\ref{prop:damping-regularizes} gives the complex-frequency description: for real $\kappa$, the resonant frequencies move below the real axis and their imaginary parts describe temporal decay. Near a simple point of the dispersion relation, the implicit-function theorem also permits a local parametrization with $\omega\in\mathbb{R}$ and complex $\kappa$, whose imaginary part describes spatial attenuation. These are locally related parametrizations of the same damped dispersion relation; no global one-to-one correspondence between their branches is asserted.
\end{remark}

\subsection{Real-frequency attenuation and the zero-damping limit}

Theorem~\ref{prop:damping-regularizes} describes damping by keeping $\kappa$ real and allowing the resonant frequency to become complex. A complementary real-frequency description keeps $\omega\in\mathbb{R}$ and continues the Bloch parameter as
\begin{equation}
    \kappa(\omega)=\kappa_r(\omega)+i\kappa_i(\omega).
\end{equation}
The imaginary part $\kappa_i$ then controls spatial decay through the lattice. Peaks of $|\kappa_i(\omega)|$ provide a directional attenuation diagnostic, but they are not complete band gaps in the self-adjoint sense: the result depends on the direction of continuation and on which complex Bloch branch is followed.

For small $\gamma>0$, the damped pole remains close to the real axis and
\begin{equation}
    |\xi(\omega)|
    \approx
    \frac{|r_\gamma|}
    {\sqrt{(\omega-\omega_0)^2+\eta^2}},
    \qquad
    \eta=\frac{\gamma}{2\beta}.
\end{equation}
Thus the contrast is finite on the real axis for every $\gamma>0$, while its maximum grows like $1/\gamma$ as $\gamma\to0^+$. The zero-damping limit is therefore singular near the material pole: weak damping preserves a recognizable remnant of the undamped resonant structure, while moving it off the real axis and producing finite temporal decay together with a frequency-dependent spatial attenuation profile. The numerical calculations in the following section display these two different representations side by side.

\section{Numerical investigations}
\label{sec:numerical-investigations}

\begin{figure}
    \centering
    \begin{tikzpicture}
        \node at (-2,-1) {\includegraphics[width=0.2\textwidth,trim=0 0 12cm 0,clip]{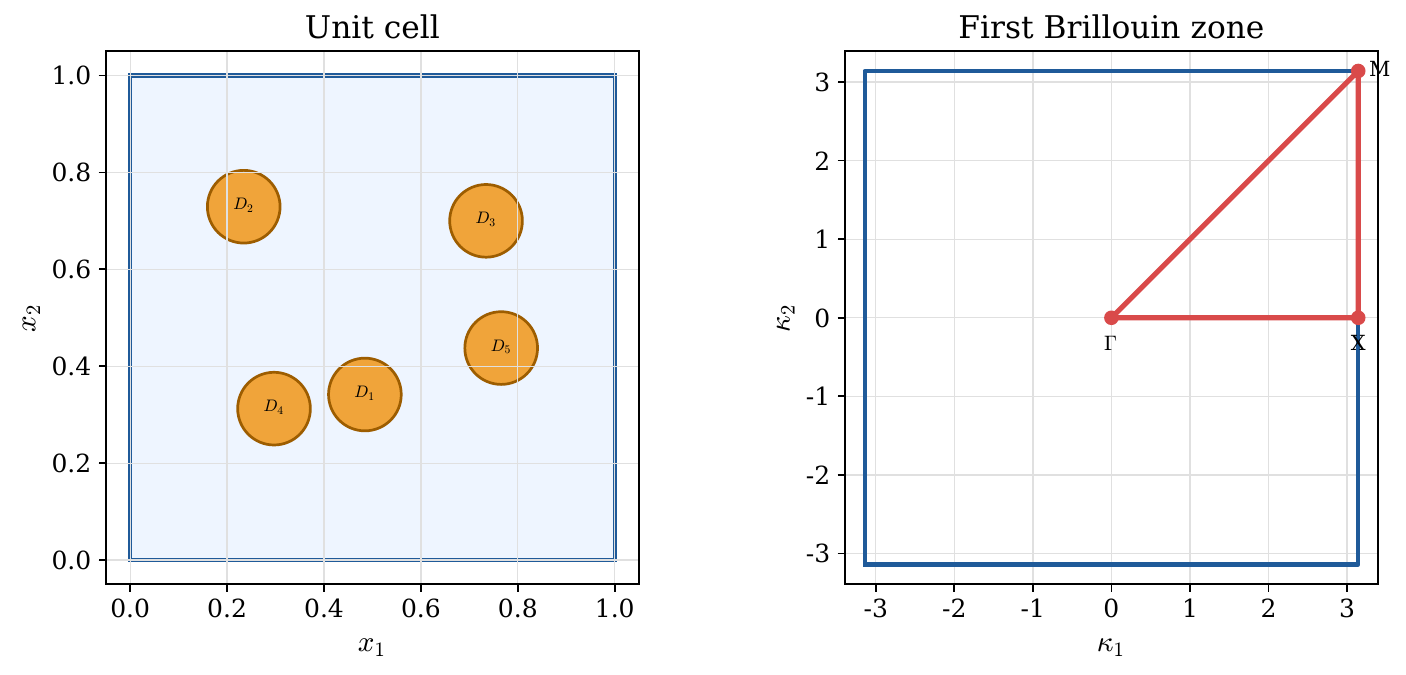}};
        \node at (1.5,-1) {\includegraphics[width=0.2\textwidth,trim=12cm 0 0 0,clip]{Plots/geometry_and_brillouin_zone.pdf}};
        \node at (8,-1) {\includegraphics[width=0.49\textwidth]{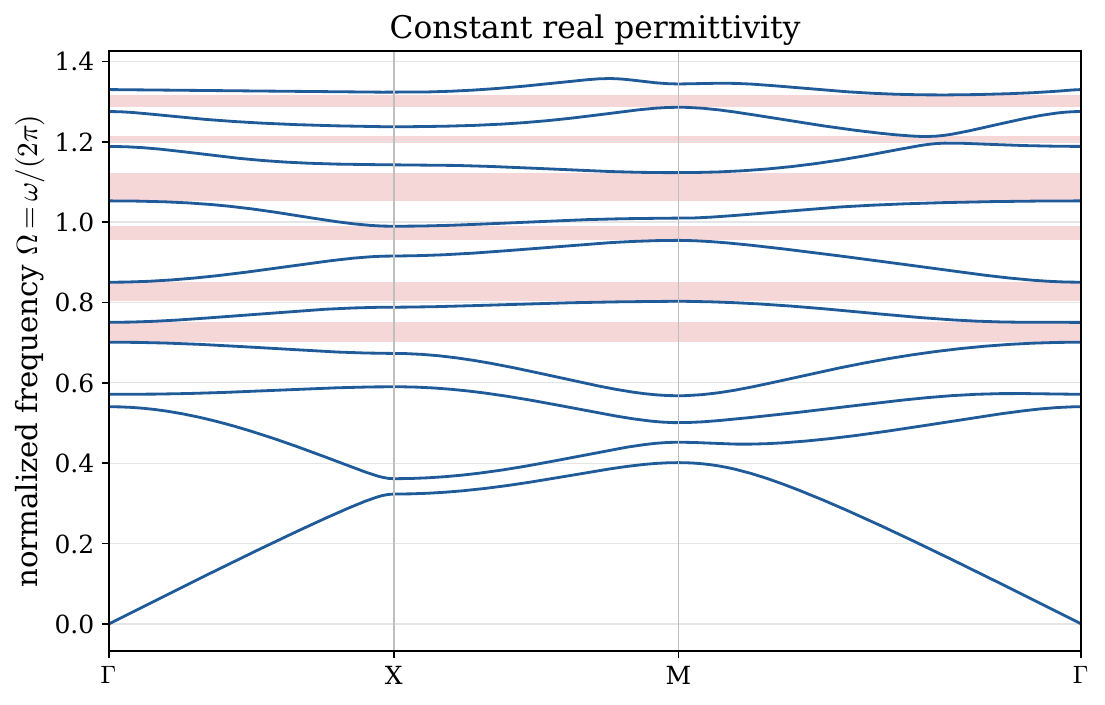}};
    \end{tikzpicture}
    \caption{Dispersion curves for constant real-valued permittivity (no dispersion or damping). Left shows the unit cell geometry (containing five circular inclusions $D_1,\dots,D_5$ arranged in a square periodic cell) and the high-symmetry path $\Gamma$--$X$--$M$--$\Gamma$ in the first Brillouin zone. The right shows the dispersion curves plotted around this path, with shaded red regions indicating gaps detected along this path. }
    \label{fig:constant-real-bands}
\end{figure}

We now illustrate the spectral mechanism identified above in a two-dimensional plane-wave model. The calculations first establish a non-dispersive reference spectrum, then exhibit the resonant cluster generated by a real material pole, and follow its regularization under damping in both the complex-frequency and complex-Bloch descriptions. A final fixed-geometry benchmark compares the asymptotic reduced solver with the plane-wave solver for selected complex-frequency branches. Figure~\ref{fig:constant-real-bands} shows the geometry used in the numerical experiments. The left panel shows the five inclusions in the unit cell, while the central panel shows the path $\Gamma$--$X$--$M$--$\Gamma$ around the first Brillouin zone. Full implementation details, numerical parameters, post-processing conventions and convergence checks are outlined in Appendix~\ref{app:numerical-methods}.

\subsection{From non-dispersive bands to a singular resonant cluster}

We first consider the non-dispersive real-valued configuration specified in Appendix~\ref{app:numerical-methods}. This gives the standard self-adjoint Bloch band structure. Figure~\ref{fig:constant-real-bands} shows the first several bands along $\Gamma$--$X$--$M$--$\Gamma$. The shaded red regions indicate frequency intervals in which no band crosses the sampled path. This provides a reference spectrum in which all frequencies are real and the usual self-adjoint interpretation of bands and gaps applies. It also separates the geometric effect of periodicity from the singular effects introduced later by frequency dispersion. Any gap observed in this configuration is therefore caused by the periodic contrast and the arrangement of the inclusions, rather than by material poles or damping.

For real-valued coefficients, the Bloch spectrum consists of real bands, and gaps arise from the periodic geometry and material contrast. No material-pole singularity is present. Figure~\ref{fig:constant-real-bands} records the conventional path diagram and provides the reference configuration against which the pole-induced and damped spectra below are interpreted.

Having isolated the effect of the periodic geometry, we introduce the undamped dispersive material law
\begin{align*}
    \varepsilon(\omega)
    =
    \varepsilon_0
    +
    \frac{\alpha}{1-\beta\omega^2},
\end{align*}
with a real material pole at $\Omega^* = 0.9$. The numerical parameter values are collected in Appendix~\ref{app:numerical-methods}. 

In Figure~\ref{fig:undamped-pole-branches}, we see the resulting Bloch branches near the real pole, together with a closer view of the branch clustering around $\Omega^*$. The red dashed horizontal line marks the material pole. This introduces a spectral feature that is absent from the non-dispersive reference problem. The branch clustering is now organized around the pole of $\varepsilon(\omega)$ and is therefore tied to the nonlinear dependence of the eigenvalue problem on frequency. The comparison with Figure~\ref{fig:constant-real-bands} isolates the singular mechanism predicted by the theory before any damping is introduced.

\begin{figure}
    \centering
    \begin{subfigure}{0.49\textwidth}
        \centering
        \includegraphics[width=\textwidth]{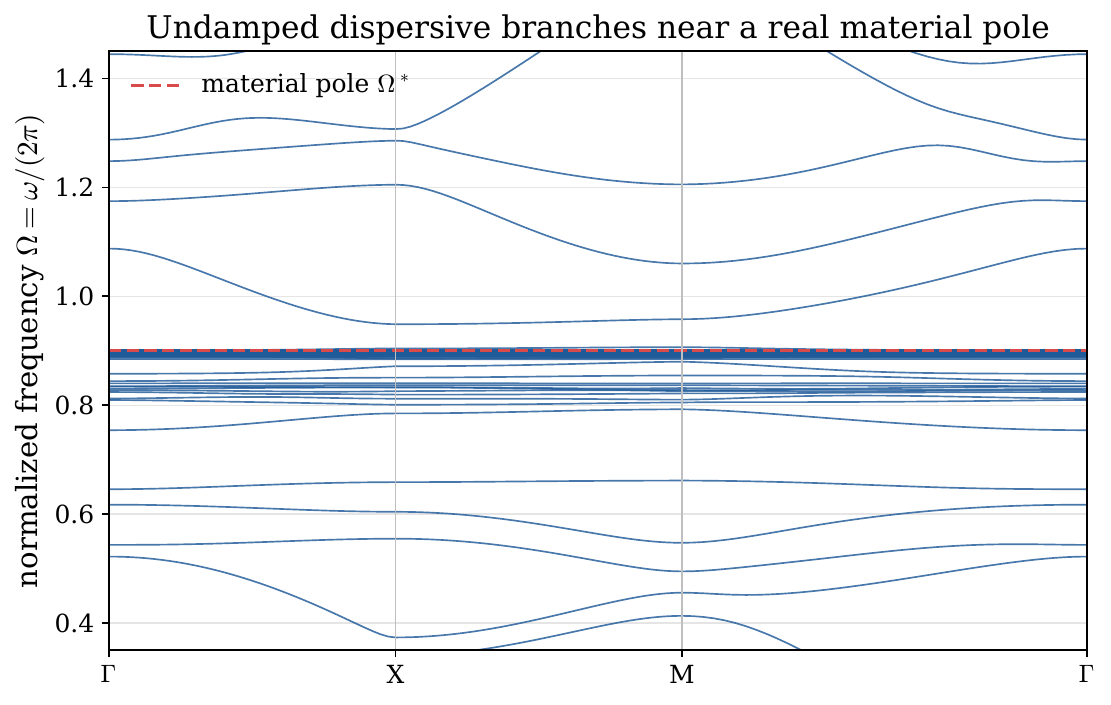}
        \caption{Branches near the pole.}
    \end{subfigure}
    \hfill
    \begin{subfigure}{0.49\textwidth}
        \centering
        \includegraphics[width=\textwidth]{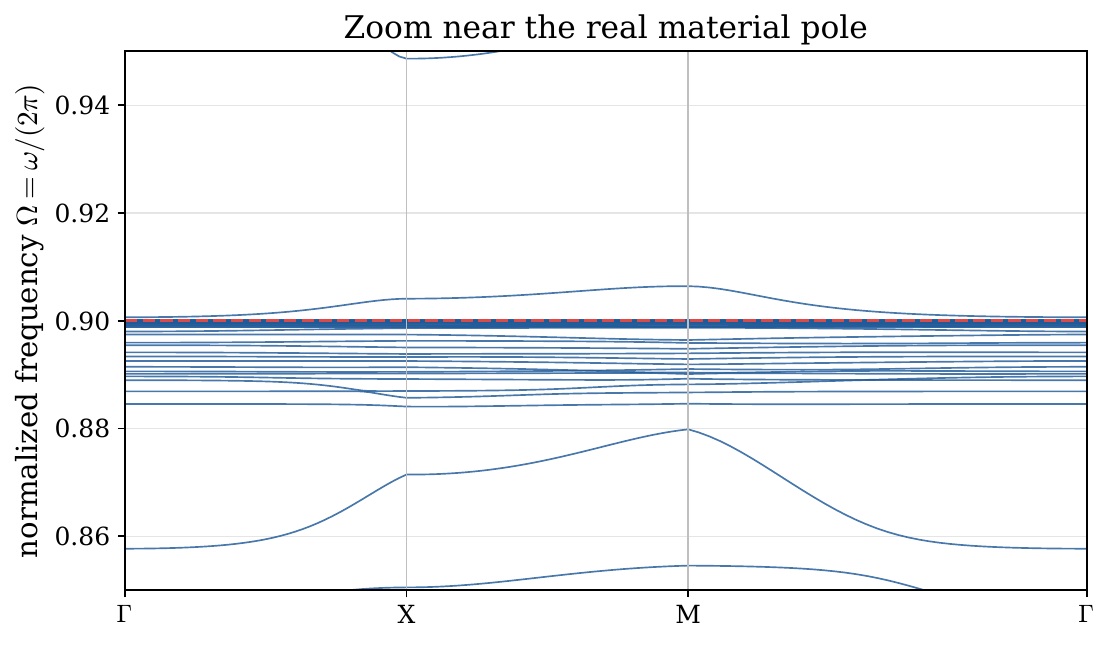}
        \caption{Zoom near $\Omega^*$.}
    \end{subfigure}
    \caption{Resonant Bloch branches for an undamped dispersive material law with a real pole at $\Omega^*=0.9$. The dashed red line marks the pole. The left panel shows the branch structure along the high-symmetry path, while the right panel zooms into the singular region. The cluster of computed branches near $\Omega^*$ illustrates how a real material pole generates a family of nearby dispersion curves.}
    \label{fig:undamped-pole-branches}
\end{figure}

The zoomed panel in Figure~\ref{fig:undamped-pole-branches} shows the numerical analogue of the singularity mechanism described in Section~\ref{sec:material-singularities}. Thus the progression from Figure~\ref{fig:constant-real-bands} to Figure~\ref{fig:undamped-pole-branches} separates the ordinary geometric band structure from the additional family of resonances organized by the material pole.

\subsection{Temporal and spatial signatures of regularization}

We next introduce material damping by taking $\gamma>0$. The material poles then move from the real axis into the lower half-plane. As explained in Section~\ref{sec:material-damping}, this removes the real-axis singularity while preserving the associated resonances as complex spectral branches. We examine this regularization using two complementary local parametrizations of the dispersion relation. First, we keep the Bloch parameter $\kappa$ real along $\Gamma$--$X$--$M$--$\Gamma$ and solve for complex resonant frequencies $\Omega$. Second, we keep the frequency real, restrict propagation to the $\Gamma$--$X$ direction, and solve for complex $\kappa_1$ with $\kappa_2=0$. The former describes temporal decay through $\operatorname{Im}\Omega$, whereas the latter describes spatial attenuation through $\operatorname{Im}\kappa_1$. For the complex-Bloch calculation, we record at each real frequency the least attenuated mode, $\min_j \left|\operatorname{Im}\kappa_{1,j}\right|$. This quantity gives the smallest spatial attenuation among the computed complex-Bloch roots and therefore identifies the least attenuated response available along $\Gamma$--$X$.

\begin{figure}
    \centering
    \includegraphics[width=\textwidth]{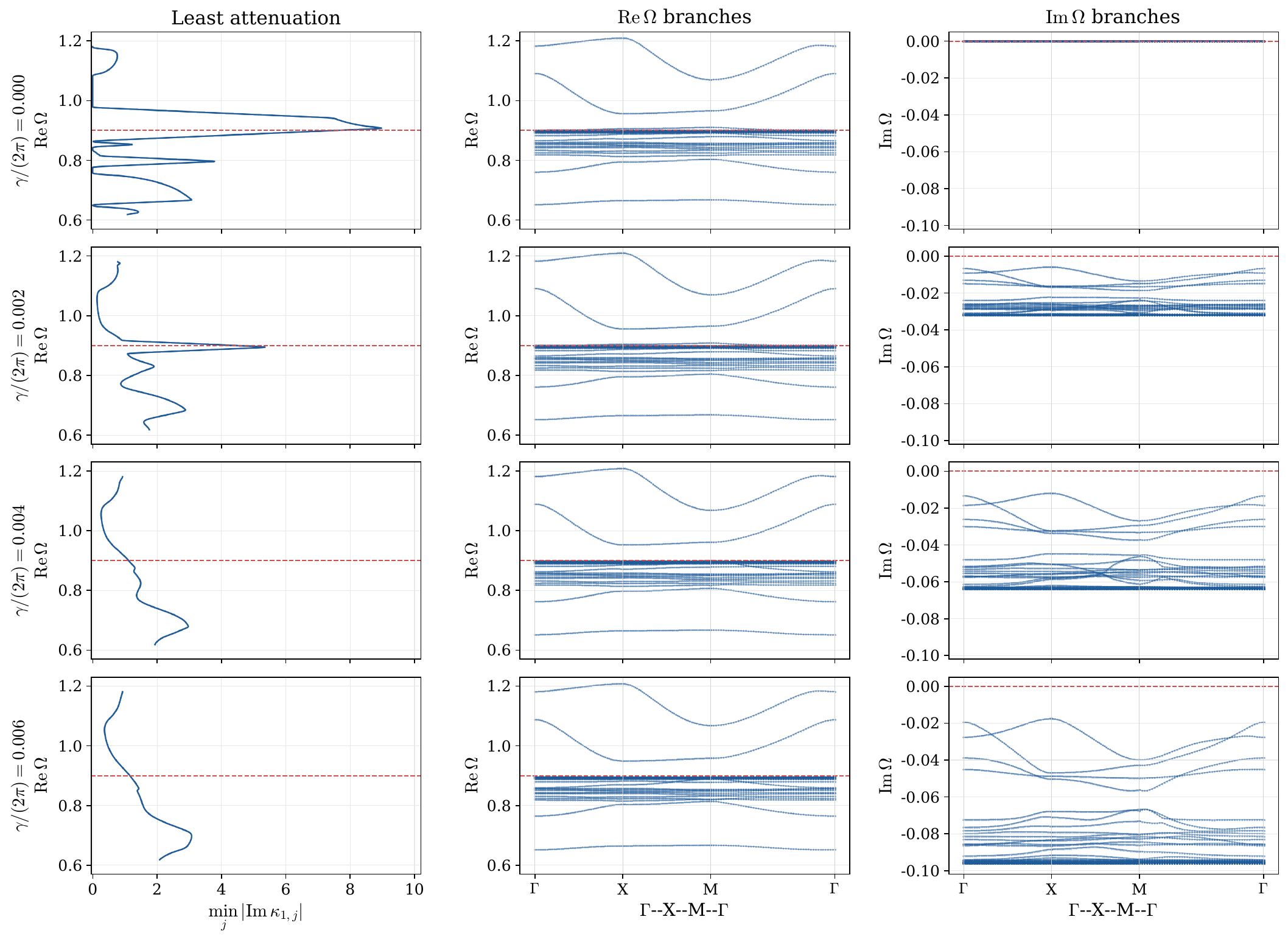}
    \caption{Temporal and spatial manifestations of increasing damping (rows). The left and middle columns show, respectively, the real and imaginary parts of the complex resonant frequencies along $\Gamma$--$X$--$M$--$\Gamma$ for real Bloch parameter. The right column shows the least attenuated complex-Bloch mode $\min_j|\operatorname{Im}\kappa_{1,j}|$ along $\Gamma$--$X$ at real frequency, with $\operatorname{Re}\Omega$ on the vertical axis to facilitate comparison with the left column. The dashed red lines mark the undamped pole frequency in the left and right columns and the real-frequency axis in the middle column. Damping displaces the resonant branches into the lower half-plane and regularizes the real-frequency attenuation profile associated with the undamped singular structure.}
    \label{fig:damping-temporal-spatial-grid}
\end{figure}

Figure~\ref{fig:damping-temporal-spatial-grid} places the two descriptions side by side at the same damping values. In the undamped row, the resonant frequencies are real and the attenuation profile contains sharp peaks near frequency intervals in which no weakly attenuated Bloch mode is available. Once damping is introduced, the resonances acquire negative imaginary parts and their projections onto the real-frequency axis no longer contain a pole. The resulting attenuation profiles are finite and vary continuously away from changes in the minimizing branch. The right column records a minimum over all computed complex-Bloch roots, so the branch realizing that minimum may change with frequency or damping. A reduction of the minimum therefore means that a less attenuated mode has become available; it does not imply that every evanescent continuation of an undamped branch has disappeared.

The complex-frequency and complex-Bloch calculations probe the same dispersion relation, but they should not be expected to agree pointwise because they use different local parametrizations and the attenuation plot minimizes over several branches. To compare their damping-induced shifts, let $(\Omega_0,\kappa_0)$ be a real undamped point on a locally analytic scalar dispersion relation
\begin{equation}
    F(\Omega,\kappa_1,\gamma)=0,
    \qquad F(\Omega_0,\kappa_0,0)=0,
\end{equation}
with the transverse Bloch component fixed. Assume $F_\Omega\neq0$ and $F_{\kappa_1}\neq0$ at this point, away from material poles and background thresholds. All derivatives below are evaluated there. The undamped branch has a real, non-zero normalized slope
\begin{equation}
    v_\Omega:=\frac{d\Omega}{d\kappa_1}
    =-\frac{F_{\kappa_1}}{F_\Omega}\in\mathbb R\setminus\{0\}.
\end{equation}
Define the temporal and spatial shifts by
\begin{equation}
    F(\Omega_0+\Delta\Omega_{\mathrm t},\kappa_0,\gamma)=0,
    \qquad
    F(\Omega_0,\kappa_0+\Delta\kappa_{\mathrm s},\gamma)=0.
\end{equation}
The implicit-function theorem gives both shifts as $O(\gamma)$. Taylor expansion then yields
\begin{align}
    F_\Omega\Delta\Omega_{\mathrm t}+F_\gamma\gamma
    &=O(\gamma^2),\\
    F_{\kappa_1}\Delta\kappa_{\mathrm s}+F_\gamma\gamma
    &=O(\gamma^2).
\end{align}
Subtracting these equations eliminates the common damping perturbation. Consequently,
\begin{equation} \label{eq:temporal-spatial-damping-shifts}
    \Delta\kappa_{\mathrm s}
    =\frac{F_\Omega}{F_{\kappa_1}}\Delta\Omega_{\mathrm t}
    +O(\gamma^2)
    =-\frac{\Delta\Omega_{\mathrm t}}{v_\Omega}+O(\gamma^2),
    \qquad
    \operatorname{Im}\Delta\kappa_{\mathrm s}
    =-\frac{\operatorname{Im}\Delta\Omega_{\mathrm t}}{v_\Omega}
    +O(\gamma^2).
\end{equation}
For a forward branch with $v_\Omega>0$, a negative leading imaginary frequency shift therefore gives positive spatial attenuation in the convention $e^{i\kappa_1x_1-i\omega t}$. The sign reverses for a backward branch, corresponding to decay in the opposite direction. This comparison concerns two different shifts induced by the same damping, not simultaneous changes of frequency and Bloch parameter along a fixed dispersion curve. It is a local weak-damping relation near a real propagating branch; it does not apply directly inside an undamped gap, at a zero-slope band edge, or through crossings and degeneracies. Nor does it identify the branch attaining the least-attenuation minimum. Figure~\ref{fig:damping-temporal-spatial-grid} therefore compares complementary spectral diagnostics rather than pointwise equivalent curves.

\subsection{Complex-plane branch motion}

\begin{figure}
    \centering
    \begin{subfigure}[c]{0.6\textwidth}
        \centering
        \includegraphics[width=\textwidth]{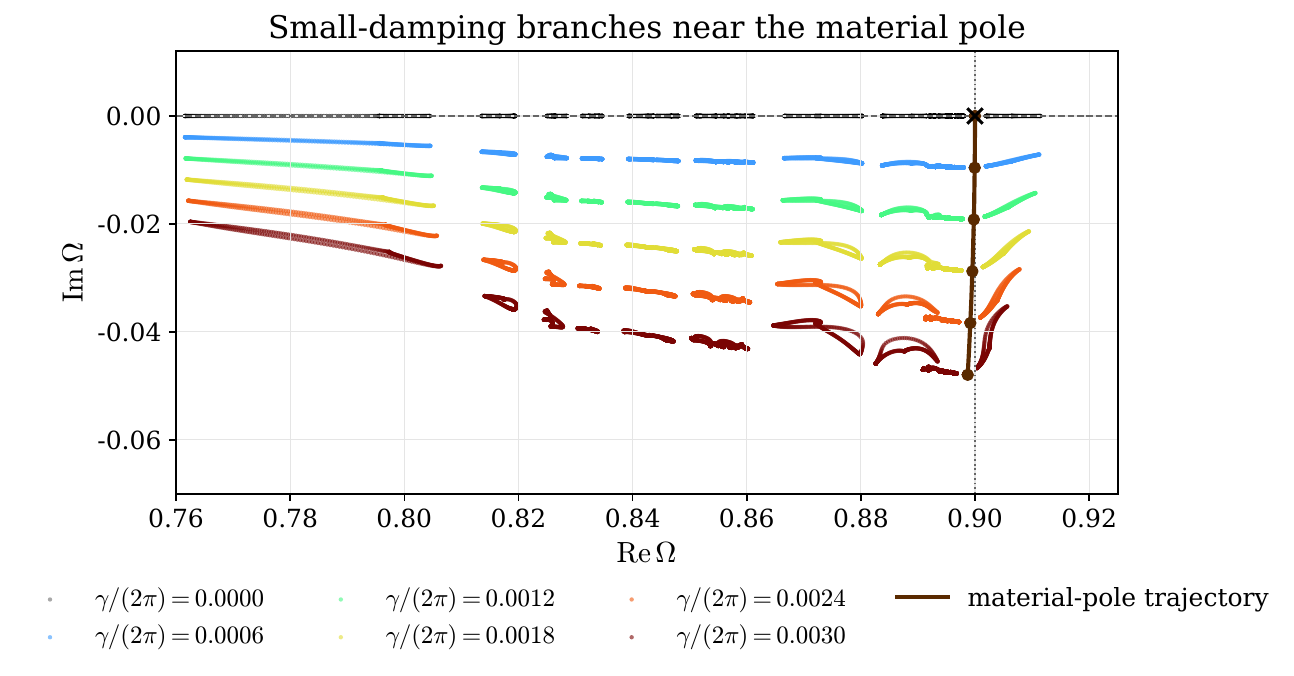}
        \caption{Resonant cluster in the complex $\Omega$-plane for small damping along $\G-X-M-\G$.}
        \label{fig:complex-omega-plane-damping}
    \end{subfigure}
    \hfill
    \begin{subfigure}[c]{0.35\textwidth}
        \centering
        \includegraphics[width=\textwidth]{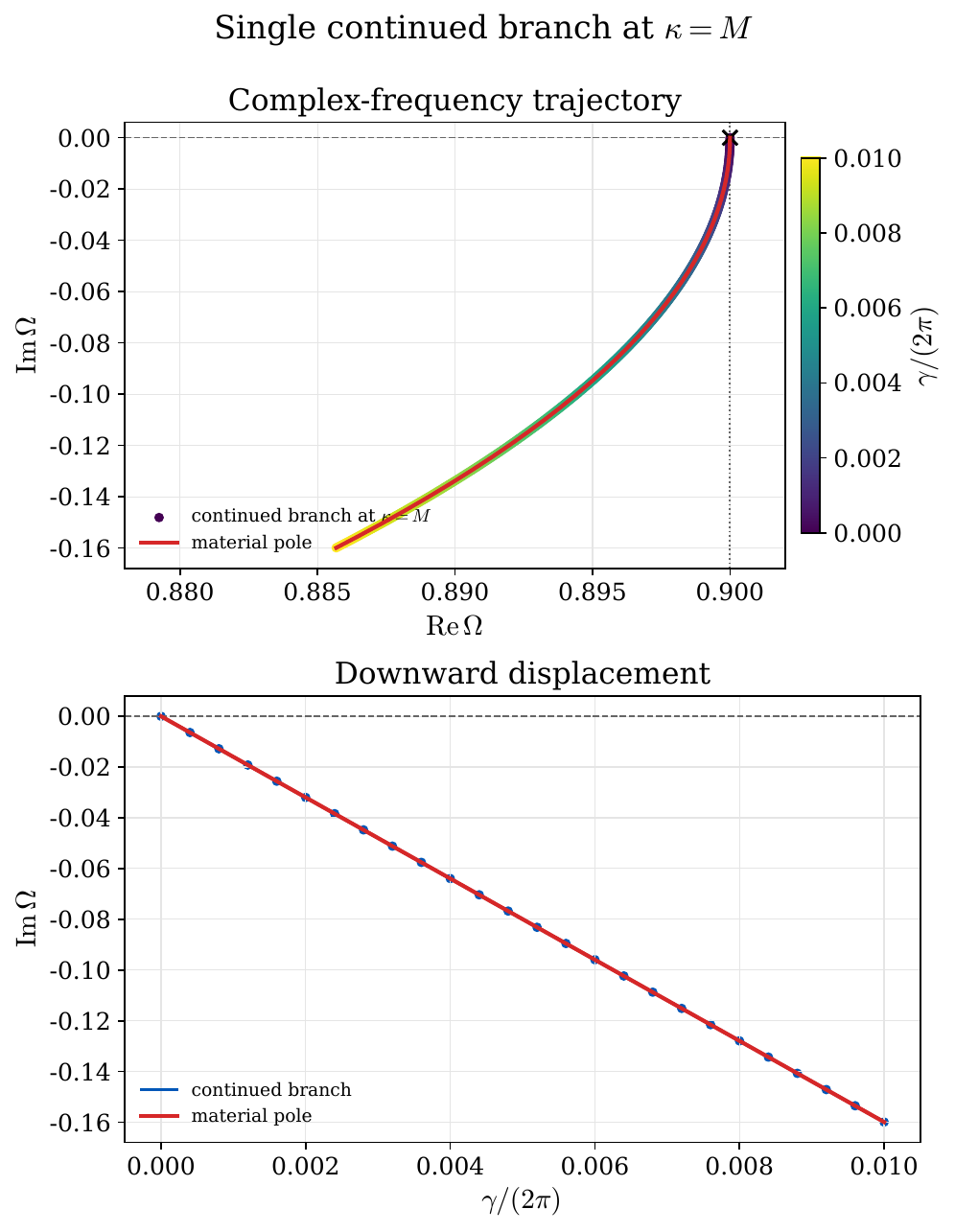}
        \caption{One tracked resonant branch at $\kappa=M$: complex trajectory (top) and imaginary part versus damping (bottom).}
        \label{fig:complex-omega-single-branch}
    \end{subfigure}
    \caption{Complex-frequency motion under damping. In (a), each colour collects resonances along the Bloch path at one damping value; the separate brown curve traces the material pole as damping varies. In (b), the tracked branch is compared with the pole trajectory and the dashed leading-order decay prediction from Theorem~\ref{prop:damping-regularizes}.}
\end{figure}

The preceding comparison shows the spectral consequences of damping in two complementary parametrizations. Let us now examine the complex-frequency motion itself, first for the resonant cluster and then for one isolated branch. Figure~\ref{fig:complex-omega-plane-damping} plots the damped resonances in the complex $\Omega$-plane near the material pole. It gives a direct numerical visualization of Theorem~\ref{prop:damping-regularizes}. The material pole moves into the lower half-plane, and the nearby resonant Bloch branches move with it. The singular real-frequency structure is thereby continued as a family of complex resonances rather than simply being erased. The focused small-damping panel also shows that the displacement is already visible in the perturbative regime, before the resonant cluster becomes strongly reorganized.

\begin{remark}[Interpretation of the complex-frequency curves]
    Figure \ref{fig:complex-omega-plane-damping} samples the closed path $\G-X-M-\G$. At fixed damping, varying $\k$ changes both the real and imaginary parts of the resonances and produces loop-like curves since the damping has broken the typical symmetries in the Brillouin zone.
\end{remark}

To make the branch-level motion explicit, we also track a single resonant branch at the fixed Bloch parameter $\kappa=M$ for small damping values. This is shown in Figure~\ref{fig:complex-omega-single-branch}. The damping range is restricted to the perturbative regime, where the branch can be continued without visible branch switching. The branch diagnostic in Figure~\ref{fig:complex-omega-single-branch} should be interpreted as a local continuation calculation, not as a global branch-tracking algorithm through the full resonant cluster. Near the material pole, many resonant branches are close together, so branch identity becomes delicate when damping is increased. Within the displayed perturbative range, the fitted imaginary displacement closely agrees with the leading-order prediction of Theorem~\ref{prop:damping-regularizes}. The quantitative comparison is reported in Appendix~\ref{app:numerical-methods}. We note that for five selected branches at fixed geometry and Bloch parameter, the reduced solver is approximately eleven times faster, with mean relative complex-frequency errors of $0.16\%$--$1.64\%$ against a refined plane-wave reference when $\delta=0.075$.

Hence, Figures~\ref{fig:damping-temporal-spatial-grid}, \ref{fig:complex-omega-plane-damping} and~\ref{fig:complex-omega-single-branch} support the regularization mechanism established in Theorem~\ref{prop:damping-regularizes}. Damping moves the resonant spectrum away from the real axis; on the real-frequency slice, its observable signature is a finite, direction-dependent spatial attenuation profile.

\section{Conclusion}

We have quantified the regularization of singular Bloch spectral branches by material damping under explicit continuation hypotheses. For a periodic array of subwavelength dispersive inclusions, the quasiperiodic Lippmann--Schwinger formulation separates the material contrast from a compact geometric operator, and a finite-dimensional reduction describes its selected modes. This makes explicit how a real pole of the permittivity causes a family of branches to collect. Their displacement from the pole is determined by eigenvalues of the geometric operator.

Damping does not eliminate these branches, it moves the material pole and the associated resonances into the lower half of the complex-frequency plane. In particular, the leading imaginary displacement is $-\gamma/(2\beta)$, independently of the Bloch parameter up to the stated subwavelength correction. The singularity on the real axis is therefore regularized by temporal decay, while the underlying resonant structure persists in the complex spectrum. The alternative real-frequency formulation reveals the corresponding spatial effect. Near a simple point of the dispersion relation, complex frequency and complex Bloch parameter are related by the local implicit-function parametrization described above, so temporal decay is accompanied by spatial attenuation. This correspondence is local: crossings, degeneracies and changes in the least-attenuated branch prevent a global pointwise identification of the two diagnostics.

The numerical calculations support each part of this picture. They display the undamped branch cluster, its displacement into the lower half-plane, and finite directional attenuation on the real-frequency slice. The isolated-branch calculation also agrees closely with the predicted leading imaginary displacement.

\section*{Acknowledgements}

The authors acknowledge the use of OpenAI's ChatGPT Codex to assist with code optimisation and literature review. The authors retain full responsibility for the mathematical proofs and arguments, and take full accountability for the accuracy and integrity of all results and conclusions presented in this work.

\appendix
\section{Operator-norm remainder estimates}
\label{app:kernel-remainders}

We justify the block expansions \eqref{eq:correct-2d-self}--\eqref{eq:correct-3d-cross}. All constants below are uniform on the compact, threshold-free parameter sets specified in Section~\ref{sec:integral-formulation}. The reference domains are bounded and the centres are fixed and distinct modulo the lattice. In three dimensions the first corrections are explicitly
\begin{align}
    \mathcal K_{ii}^{\kappa,\omega,(1)}[\varphi](X)
    &=-R_3^{\kappa,k_0}(0)\int_{D_i}\varphi(Y)\,\mathrm dY,\\
    \mathcal K_{ij}^{\kappa,\omega,(1)}[\varphi](X)
    &=-G^{\kappa,k_0}(z_i-z_j)\int_{D_j}\varphi(Y)\,\mathrm dY,
    \qquad i\neq j.
\end{align}
For the diagonal blocks, the error kernel after subtraction of the displayed leading terms is
\begin{equation}
    e_{ii}^{(d)}(X,Y)
    =-\delta^{d-2}\big[R_d^{\kappa,k_0}(\delta(X-Y))-R_d^{\kappa,k_0}(0)\big].
\end{equation}
By \eqref{eq:green-remainder-lipschitz},
\begin{equation}
    |e_{ii}^{(d)}(X,Y)|\leq C\delta^{d-1}|X-Y|.
\end{equation}
If $E_{ii}^{(d)}$ denotes the corresponding integral operator, the Hilbert--Schmidt estimate yields
\begin{equation}
    \|E_{ii}^{(d)}\|_{L^2(D_i)\to L^2(D_i)}
    \leq\left(\int_{D_i}\int_{D_i}|e_{ii}^{(d)}(X,Y)|^2\,\mathrm dY\,\mathrm dX\right)^{1/2}
    \leq C\delta^{d-1}\left(\int_{D_i}\int_{D_i}|X-Y|^2\,\mathrm dY\,\mathrm dX\right)^{1/2}.
\end{equation}
The last factor is finite, giving $O(\delta)$ in two dimensions and $O(\delta^2)$ in three dimensions.

For $i\neq j$, put $a_{ij}=z_i-z_j$. Since $a_{ij}\notin\Lambda$, for sufficiently small $\delta$ the segments $a_{ij}+t\delta(X-Y)$, $0\leq t\leq1$, stay uniformly away from every lattice singularity. Taylor's formula in integral form gives
\begin{equation}
    G^{\kappa,k_0}(a_{ij}+\delta(X-Y))-G^{\kappa,k_0}(a_{ij})
    =\delta\int_0^1\nabla G^{\kappa,k_0}(a_{ij}+t\delta(X-Y))\cdot(X-Y)\,\mathrm dt.
\end{equation}
Smoothness away from the lattice implies a uniform bound on the gradient. Thus the off-diagonal error kernel satisfies
\begin{align}
    e_{ij}^{(d)}(X,Y)
    &:=-\delta^{d-2}\big[G^{\kappa,k_0}(a_{ij}+\delta(X-Y))-G^{\kappa,k_0}(a_{ij})\big],\\
    |e_{ij}^{(d)}(X,Y)|&\leq C\delta^{d-1}|X-Y|.
\end{align}
Consequently,
\begin{equation}
    \|E_{ij}^{(d)}\|_{L^2(D_j)\to L^2(D_i)}
    \leq C\delta^{d-1}\left(\int_{D_i}\int_{D_j}|X-Y|^2\,\mathrm dY\,\mathrm dX\right)^{1/2}
    =O(\delta^{d-1}).
\end{equation}
These estimates prove all four claimed remainder orders without assuming smoothness of the Laplace-subtracted self-interaction kernels at the origin.

\section{Numerical methods}
\label{app:numerical-methods}

The spectral illustrations use five discs of radius $\delta=0.075$ in a unit square, with $\varepsilon_0=\mu_0=1$ and centres $z_1=(0.484375,0.341667)$, $z_2=(0.234375,0.729167)$, $z_3=(0.734375,0.700000)$, $z_4=(0.296875,0.312500)$ and $z_5=(0.765625,0.437500)$. We use $\Omega=\omega/(2\pi)$, $Y^*=[-\pi,\pi]^2$, non-dispersive inclusion permittivity $12$, and dispersive parameters $\alpha=8$, $\Omega^*=0.9$, $\beta=(2\pi\Omega^*)^{-2}$. Figure legends report $\gamma/(2\pi)$ and the equations below use angular frequency and $\gamma$.

\subsection{Plane-wave formulation}

For the inclusion indicator $\chi$ and contrast $\Delta\varepsilon=\varepsilon-\varepsilon_0$, use
\begin{equation}\label{eq:appendix-plane-wave-basis}
 u_\kappa^{(p)}(x)=\sum_{n\in\mathcal I_p}c_n e^{i(\kappa+G_n)\cdot x},
 \qquad G_n=2\pi n,\qquad
 \mathcal I_p=\{n\in\mathbb Z^2:|n_1|,|n_2|\leq p\}.
\end{equation}
The convolution matrix $C_{mn}=\widehat\chi_{m-n}$ is assembled from
\begin{equation}\label{eq:appendix-fourier-indicator}
 \widehat\chi_\ell=\int_Y\chi(x)e^{-2\pi i\ell\cdot x}\,\mathrm dx,
\end{equation}
using midpoint quadrature on a $220\times220$ raster. The field coefficients satisfy
\begin{equation}\label{eq:appendix-pwe-system}
 A_\kappa\mathbf c=\omega^2[\varepsilon_0I+\Delta\varepsilon(\omega)C]\mathbf c,
 \qquad (A_\kappa)_{mn}=\delta_{mn}\sum_{r=1}^2(\kappa_r+G_{n,r})^2,
\end{equation}
with algebraic squares for complex $\kappa$. For $\Delta\varepsilon=11$, this is a Hermitian generalized eigenproblem in $\omega^2$.

With $d_\gamma(\omega)=1-\beta\omega^2-i\gamma\omega$, the dispersive system is
\begin{equation}\label{eq:appendix-dispersive-polynomial}
 [d_\gamma(\omega)(A_\kappa-\varepsilon_0\omega^2I)-\alpha\omega^2C]\mathbf c=0,
\end{equation}
equivalent to the rational problem away from material poles. Damped frequencies are obtained directly, retaining $\operatorname{Re}\omega>0$ and $\operatorname{Im}\omega\leq0$ up to tolerance. The complex-frequency panels of Figure~\ref{fig:damping-temporal-spatial-grid} use $p=3$ and $50$ subdivisions per path segment.

At real $\omega$ and $\kappa_2=0$, solve instead
\begin{equation}\label{eq:appendix-complex-kappa-polynomial}
 [\kappa_1^2I+2\kappa_1G_x+G_x^2+G_y^2
 -\omega^2(\varepsilon_0I+\Delta\varepsilon(\omega)C)]\mathbf c=0,
\end{equation}
where $G_x=\operatorname{diag}(G_{n,1})$ and $G_y=\operatorname{diag}(G_{n,2})$. At each frequency, independently, the raw attenuation is the minimum $|\operatorname{Im}\kappa_1|$ over finite roots with $-10^{-6}\leq\operatorname{Re}\kappa_1\leq\pi+10^{-6}$, or over all finite roots if this set is empty.

\subsection{Figure construction}

For the complex-frequency panels of Figure~\ref{fig:damping-temporal-spatial-grid}, one-to-one nearest-neighbour matching uses a maximum step $0.075$ in $\Omega$. Tracks covering less than $96\%$ of the path are omitted, as are tracks with component steps above $0.045$ (real part) or $0.018$ (imaginary part); the two panels may therefore retain different subsets.

The attenuation profiles apply outlier detection, interpolation across flagged jump neighbourhoods, median and moving-average filtering, and a pointwise change limiter; residual isolated spikes are locally interpolated. These operations modify sampled profiles, not eigenpairs, and are distinct from branch continuation. Figure~\ref{fig:complex-omega-single-branch} instead uses a predictor based on the preceding root and material-pole displacement, without attenuation-profile filtering; the fitted slope of $\operatorname{Im}\Omega$ against $\gamma/(2\pi)$ is compared with $-1/(2\beta)$.

\subsection{Fixed-geometry damping benchmark}
\label{app:damping-benchmark}

Figure~\ref{fig:damping-solver-benchmark} uses $\delta=0.075$, $\kappa=M$ and $g=\gamma/(2\pi)=0,0.002,\ldots,0.06$. One constant mode per disc gives
\begin{align}
 (\mathsf M^{\mathrm{asy}}_\kappa)_{ii}
 &=-\tfrac12\log\delta+\tfrac18-\pi R_2^{\kappa,\omega}(0),
 &(\mathsf M^{\mathrm{asy}}_\kappa)_{ij}
 &=-\pi G^{\kappa,\omega}(z_i-z_j),\quad i\ne j
 \end{align}
 and
 \begin{align}
 \det[d_\gamma(\omega)I_5-\delta^2\alpha\omega^2
 \mathsf M^{\mathrm{asy}}_\kappa(\omega,\delta)]=0.
\end{align}
Undamped roots are bracketed between diffraction thresholds (scan spacing $0.003$); damping continuation uses predictor--corrector steps of at most $0.002$, with step halving and overlap threshold $0.9$.

The benchmark uses analytic Fourier coefficients
\begin{equation}
 \widehat\chi_\ell=\pi\delta^2
 \frac{2J_1(2\pi\delta|\ell|)}{2\pi\delta|\ell|}
 \sum_{j=1}^5e^{-2\pi i\ell\cdot z_j},
\end{equation}
with quotient $1$ at $\ell=0$.

\subsection{Cost and frequency comparison at fixed resonator size}
\label{subsec:damping-efficiency}

\begin{figure}
    \centering
    \begin{subfigure}{0.49\textwidth}
        \centering
        \includegraphics[width=\textwidth]{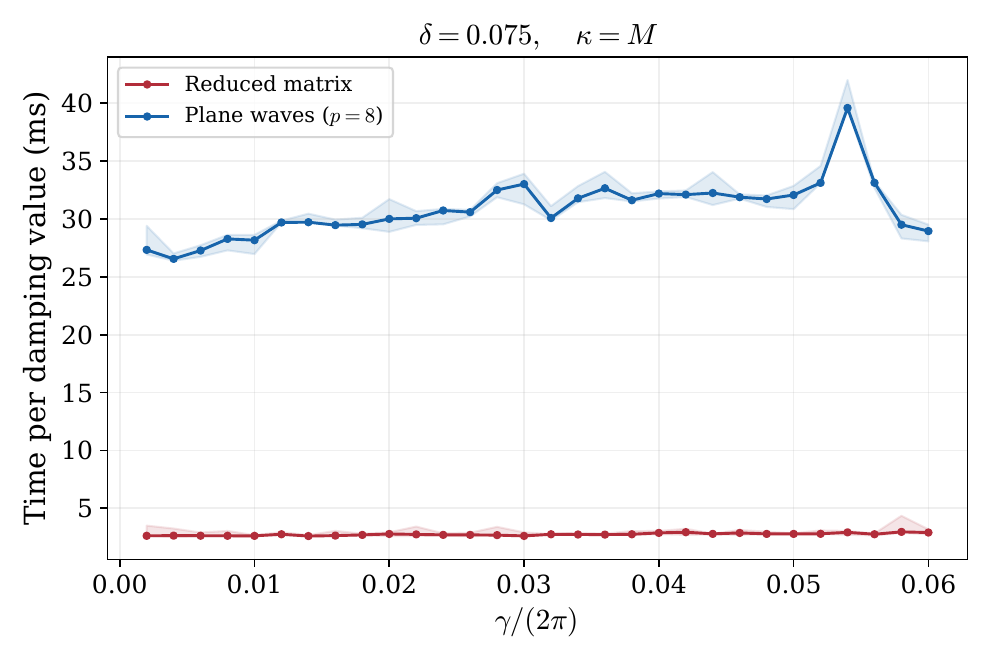}
        \caption{Runtime per damping value.}
        \label{fig:damping-solver-runtime}
    \end{subfigure}
    \hfill
    \begin{subfigure}{0.49\textwidth}
        \centering
        \includegraphics[width=\textwidth]{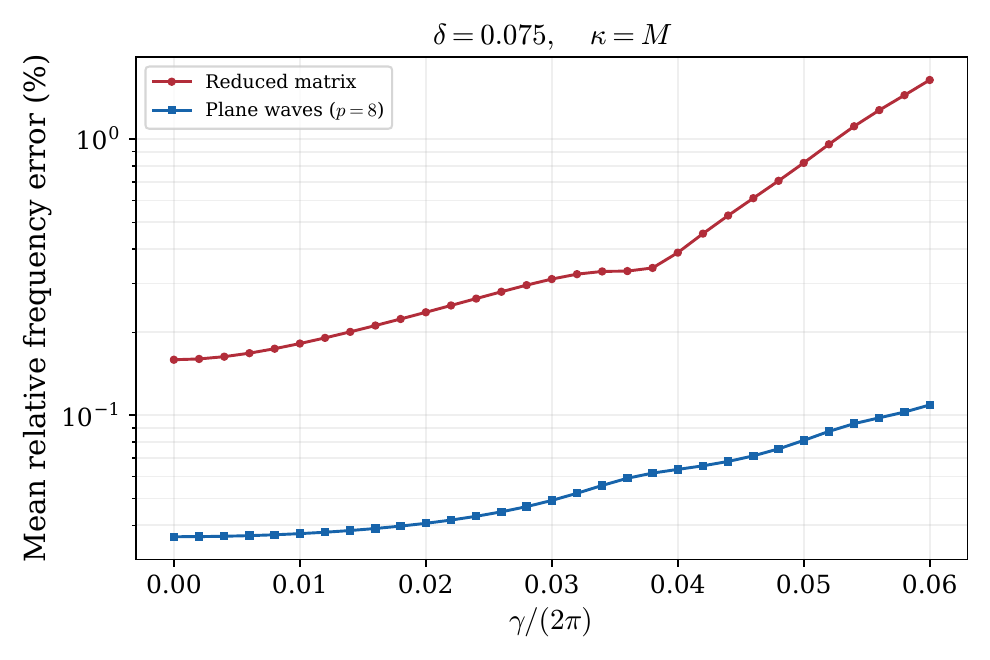}
        \caption{Mean relative frequency error.}
        \label{fig:damping-solver-accuracy}
    \end{subfigure}
    \caption{Fixed-geometry comparison at $\delta=0.075$ and $\kappa=M$. Left: median runtime for five selected modes at each positive damping value; shading gives the range over three repetitions. Geometry assembly and zero-damping initialization are excluded. Right: arithmetic mean of the relative complex-frequency errors against the $p=20$ plane-wave reference, including zero damping.}
    \label{fig:damping-solver-benchmark}
\end{figure}

Figure~\ref{fig:damping-solver-benchmark} compares the reduced interaction-matrix solver described above with the full plane-wave formulation at fixed radius $\delta=0.075$, the same five inclusion centres and unit lattice period as in the preceding illustrations, and $\kappa=M$. We follow five selected modes over $g=\gamma/(2\pi)=0,0.002,\ldots,0.06$.

The reference resonances are computed using the same plane-wave formulation as the timed full solver, but with a larger Fourier cutoff: $p=8$ ($289$ Fourier modes) for the timed calculation and $p=20$ ($1681$ modes) for the reference. At each resolution, we solve \eqref{eq:appendix-dispersive-polynomial}, select the five lowest positive undamped frequencies, and continue the corresponding modes using frequency prediction and eigenvector matching. The reference frequencies $\Omega_j^{\mathrm{ref}}(g)$ are the resulting $p=20$ roots. Thus the plane-wave error measures finite-cutoff differences, whereas the reduced-solver error also includes the asymptotic modal approximation. 

For either solver $S$, the plotted mean relative complex-frequency error is
\begin{equation}
    E_S(g)=\frac15\sum_{j=1}^5
    \frac{|\Omega_j^S(g)-\Omega_j^{\mathrm{ref}}(g)|}
         {|\Omega_j^{\mathrm{ref}}(g)|}.
\end{equation}
Reference refinement from $p=16$ to $20$ changes individual frequencies by less than $1.22\times10^{-4}$ relatively. Reducing continuation steps to $0.0005$ changes reduced roots by less than $1.8\times10^{-11}$.

Timings are medians of three single-threaded runs, including mode selection but excluding geometry assembly, zero-damping initialization, reference calculations, validation and plotting. Initialization costs $12.3$ ms (reduced) and $27.5$ ms (full). The ratio of median summed positive-damping solve times is $10.95$, or $9.83$ including initialization.

The reduced solver is faster at every sampled positive damping value, with an approximately elevenfold reduction in the summed solve time for the tested configurations. Its mean relative frequency error increases from approximately $0.16\%$ to $1.64\%$, whereas the timed plane-wave errors range from approximately $0.036\%$ to $0.109\%$. These averages concern five selected modes at fixed $M$, not the full spectrum or Brillouin zone. The reduced solver therefore provides inexpensive approximations, while the full discretization is more accurate relative to the refined reference.

\bibliographystyle{abbrv}
\bibliography{references}{}

\end{document}